\documentclass[10pt,a4paper]{article}

\usepackage{jheppub}

\usepackage{amsmath}
\usepackage{amssymb}
\usepackage{graphicx,color}

\usepackage{amsmath,amsthm}
\usepackage{amssymb}

\usepackage{graphicx}
\usepackage{bm}

\newcommand{\Ref}[1]{(\ref{#1})}
\newtheorem{Definition}{Definition}[section]
\newcommand{\half}{\frac{1}{2}}
\newcommand{\Uqsl}{\mathrm{U}_q(\fs\fl_2)}
\newcommand{\slq}{\mathrm{SL}_q(2)}

\newcommand{\su}{\fs\fu_2}
\newcommand{\suq}{\mathrm{SU}_q(2)}
\newcommand{\Uqsu}{\mathrm{U}_q(\fs\fu_2)}

\def\be{\begin{eqnarray}}
\def\ee{\end{eqnarray}}

\newcommand{\ca}{\mathcal A}

\newcommand{\cf}{\mathcal F}

\newcommand{\cp}{\mathcal P}

\newcommand{\calr}{\mathcal R}
\newcommand{\cs}{\mathcal S}
\newcommand{\ct}{\mathcal T}

\newcommand{\fl}{\mathfrak{l}}

\newcommand{\fs}{\mathfrak{s}}  
  
\newcommand{\fu}{\mathfrak{u}}

  \newcommand{\Fx}{\mathfrak{X}}

\renewcommand{\a}{\alpha}
\renewcommand{\b}{\beta}
\newcommand{\g}{\gamma}
\newcommand{\G}{\Gamma}

\newcommand{\eps}{\varepsilon}

\newcommand{\sig}{\sigma}

\renewcommand{\l}{\lambda}

\renewcommand{\o}{\omega}

\renewcommand{\t}{\tau}

\newcommand{\rmd}{\mathrm d}

\newcommand{\lt}{\left}
\newcommand{\rt}{\right}

\newcommand{\lag}{\left\langle}
\newcommand{\rag}{\right\rangle}

\newcommand{\tr}{\mathrm{tr}}
\newcommand{\bbc}{\mathbb{C}}

\newcommand{\id}{\mathrm{id}}
\newcommand{\Pol}{\mathrm{Pol}}

\newcommand{\Rhd}{\rhd\!\!\!\!\!{}^{^R}\,}
\newcommand{\Lhd}{\rhd\!\!\!\!\!{}^{^L}\,}

\title{On the Asymptotics of Quantum Group Spinfoam Model}

\author[a]{You Ding}         \author[a]{          Muxin Han}

\affiliation[a]{Centre de Physique Th\'eorique%
\footnote{Unit\'e mixte de recherche (UMR 6207) du CNRS et des Universit\'es de Provence (Aix-Marseille I), de la Meditarran\'ee (Aix-Marseille II) et du Sud (Toulon-Var); laboratoire affili\'e \`a la FRUMAM (FR 2291).}, CNRS-Luminy Case 907,  F-13288 Marseille, France}

\abstract{Recently a quantum group deformation of EPRL spinfoam model was proposed in \cite{EPRLq} by one of the authors, and in \cite{EPRLq2} by Fairbairn and Meusburger. It is interesting to study the high spin asymptotics of the quantum group spinfoam model, to see if it gives the discrete Einstein gravity with cosmological constant as its semiclassical limit. In this article we propose a new technique, which can simplify the analysis of the high spin asymptotics for quantum group spinfoam vertex amplitude. This technique can generalize the spinfoam asymptotic analysis developed by Barrett, et al to quantum group spinfoam. As a preparation of asymptotic analysis, we define and analyze the coherent states and coherent intertwiners for quantum group, which has certain ``factorization properties''. We show that in the high spin limit of quantum group spinfoam, many q-deformed noncommutative ingredients become classical and commutative. In particular, the squared norm of coherent intertwiner and the (Euclidean) vertex amplitude become integrals on classical group, while there are some additional terms (written in terms of classical group variables) make quantum group corrections to the usual (classical group) coherent intertwiner and vertex amplitude. These quantum group correction terms turn out to be proportional to the deformation parameter, which hopefully gives the cosmological term as its semiclassical limit.

}


\arxivnumber{}

\begin{document}

\maketitle

\section{Introduction}

The infra-red divergences of spinfoam models are often regularized by the deformation of spinfoam model from a classical group to a quantum group (see \cite{kassel,klimyk} for an introduction of quantum group), while the quantum group spinfoam models are expected to give the discrete gravity with cosmological constant as its semiclassical limit. For 3-dimensional gravity, the Turaev-Viro model \cite{TV} is a deformation of the Ponzano-Regge model \cite{PR} by the quantum group $\Uqsu$ ($q$ is a root of unity). The partition function of the Turaev-Viro model are finite and defines some invariants of 3-manifolds, because there is a quantum group cut-off of the admissible spin $j_f$ on each face. Moreover, the semiclassical limit of Turaev-Viro amplitude gives the 3-dimensional Regge action with a cosmological constant \cite{MT}. In 4-dimensions, the Crane-Yetter spin-foam model \cite{CY} is a deformation of 4-dimensional SU(2) BF theory (the Ooguri model \cite{Ooguri}) by $\Uqsu$ ($q$ is a root of unity). Similar to 3-dimensional case, the partition function of the Crane-Yetter model is finite and defines a topological invariant of 4-manifolds \cite{CKY}. Moreover the Crane-Yetter partition function is also a partition function of 4-dimensional SU(2) BF theory with a cosmological constant.

The lessons from 3-dimensional gravity and 4-dimensional topological field theory suggest that the quantum group deformation of 4-dimensional spinfoam models for quantum gravity gives a finite partition function, which can also be considered as a spinfoam model for quantum gravity with a cosmological constant. Such an idea was first suggested in \cite{smolin}, where the $\Uqsu$ spin-networks and its geometric operators were studied. Then in \cite{BCq}, a quantum-group-deformed Barrett-Crane model was proposed, and was shown that the resulting spinfoam partition function is finite. Recently the quantum group deformation of EPRL spinfoam model was proposed in \cite{EPRLq} (Lorentizian case) by one of the authors, and in \cite{EPRLq2} (Lorentzian \& Euclidean) independently by Fairbairn and Meusburger, where it was shown that such a quantum group deformation also gives a \emph{finite} spinfoam partition function. Then the next task is to study the high spin limit of the quantum group spinfoam model, in order to see if it gives the discrete Einstein gravity with a cosmological constant.

The present article contributes as a first step toward the high spin asymptotic analysis of the quantum group spinfoam model. The conventional high spin analysis for the Turaev-Viro model uses the asymptotic behavior of the quantum 6j symbol, which is obtained by its recursion relation (see \cite{MT}). But it is difficult to generalize this technique to 4-dimensional spinfoam models. Here, however from the lessons we learned from the asymptotic analysis for classical group EPRL spinfoam vertex in \cite{semiclassical} developed by Barrett et al, we propose a new technique to analyze the high spin asymptotics for quantum group spinfoam vertex amplitude, which can generalize the method in \cite{semiclassical} to quantum group spinfoam model.

In order to generalize the high spin asymptotic analysis in \cite{semiclassical} to quantum group case, a first step is to generalize the notion of Perelomov's coherent state to quantum group. Such a generalization was developed in \cite{JS} for general compact quantum group, and in e.g. \cite{suqcoherent} for $\Uqsu$. However here we will present an independent derivation for the coherent state on the quantum group $\slq$ by using the $\slq$-comodule-algebra and quantum plane\footnote{Actually the coherent state we define here is slightly different from the one introduced in \cite{suqcoherent}. we can see the difference from their over-completeness relations.}. Within this derivation we show a manifest ``factorization property'' for the quantum group coherent state, which is crucial for our analysis of high spin asymptotics.

However once we look into the properties of quantum group coherent state, we immediately see a potential mismatching problem: From the asymptotic analysis in \cite{semiclassical}, an SU(2) coherent state $|j,\mathbf{n}\rangle$ corresponds to a vector $j\hat{n}$, which is a face area vector of a classical tetrahedron in the classical (commutative) boundary geometry. However in \cite{suqcoherent}, it was shown that the $\Uqsu$ coherent state peaks at the geometry of a q-sphere \cite{qsphere}, which has non-commutative structure. What we want to find from the quantum group spinfoam model is a classical gravity with cosmological constant. Thus the boundary geometry must be a classical commutative geometry, although it is curved by the cosmological constant. Therefore conceptually, we have a potential mismatching problem between a noncommutative q-geometry and a classical curved geometry.

The noncommutativity of quantum group leads to another possible technical problem for the asymptotic analysis: For the usual EPRL model with classical group, in \cite{semiclassical} the asymptotic analysis was done by using the stationary phase approximation, since the vertex amplitude there can be written as a Haar integration over classical group (Spin(4) or SL(2,$\bbc$)). However in the case of quantum group, although the vertex amplitude can be written as a \emph{quantum} Haar integral, the integrand is a noncommutative function on quantum group, and moreover, the quantum group Haar integral can only be understood as a linear functional on the space of noncommutative functions on quantum group, rather than a usual integral. Therefore we have a technical problem about how to make the stationary phase approximation in the context of noncommutative structure from quantum group.

In the following we develop an approximation scheme for the high spin limit to solve both of the above two problems. We find that fortunately, in the high spin limit of the quantum group spinfoam model, lots of noncommutative quantum group ingredients become classical and commutative in a certain sense. Take the Euclidean quantum group vertex amplitude\footnote{In the following, we define an Euclidean spinfoam vertex amplitude using the quantum group $\slq$ with $q$ an arbitrary complex number. However in our opinion, if we restrict ourself to the case of $q$ a root of unity, the Euclidean vertex amplitude defined here would coincide with the one proposed in \cite{EPRLq2}.} as an example, we can make a certain power expansion with respect to the quantum group deformation parameter $\o$, where we find that the leading term is nothing but the contribution from the classical group spinfoam vertex, and there is also a full tower of quantum group corrections. Moreover in the high spin limit:
\begin{enumerate}
  \item The leading term and all the quantum group corrections become the \emph{commutative} functions on \emph{classical} group.
  \item The quantum Haar integral recovers the usual Haar integral on \emph{classical} group. 
  \item The leading contribution of the quantum group corrections in the high spin limit is proportional to the deformation parameter $\o$, which is the \emph{lowest} order correction from the quantum group.
\end{enumerate}
Finally it turns out that the resulting high spin asymptotics coincide with the following classical integration (an effective vertex amplitude)
\be
\ca_{v,\text{eff}}=\int\rmd g\ e^{\cs_{cl}+\o\times(\text{correction})}.
\ee
If we take $\o=0$, the above formula becomes $\int\rmd g\ e^{\cs_{cl}}$, which is the usual (classical group) EPRL spinfoam vertex amplitude. Further research should be carried out on the stationary phase analysis of the above integral, where the quantum group correction should deform the stationary point for the classical group EPRL spinfoam vertex, which is expected to make the 4-simplex curved.

\section{Quantum Group and Coherent State}

\subsection{Quantum Groups $\slq$ and $\suq$}

First of all, we introduce some conventions and notations. we define a complex deformation parameter $q=e^{-\o}\in\bbc$. The quantum groups recover the corresponding classical groups as $q\to1$ or $\o\to0$. Given an complex number $z\in\mathbb{C}$, the deformed q-number $[z]$ is defined by
\be
[z]=\frac{q^z-q^{-z}}{q-q^{-1}}
\ee
As $q\to1$, the deformed number recovers its classical limit, i.e. $\lim_{q\to1}[z]=z$. For any non-negative integer $n$, we can define the deformed factorials
\be
[0]!=1\ \ \ \ \ [n]!:=[1][2]\cdots[n]
\ee

%We recall the definition of the quantum group $\Uqsu$ and review the basic facts about $\Uqsu$ (see \cite{kassel} for details).

%\begin{Definition}

The Hopf algebra $\Uqsl$ is an universal enveloping algebra generated by four generators $q^{\pm J_z}$ and  $J_\pm$ with the algebraic relations
\be
q^{\pm J_z}q^{\mp J_z}=1\ \ \ \ \ q^{J_z}J_\pm q^{-J_z}=q^{\pm1}J_\pm\ \ \ \ \ \lt[J_+,\ J_-\rt]=\frac{q^{2J_z}-q^{-2J_z}}{q-q^{-1}}\label{qSU}
\ee
The comultiplication $\Delta:\ \Uqsl\to\Uqsl\otimes\Uqsl$, as an algebra morphism, is defined by
\be
\Delta(q^{\pm J_z})=q^{\pm J_z}\otimes q^{\pm J_z}\ \ \ \ \ \Delta(J_\pm)=q^{-J_z}\otimes J_\pm+J_\pm\otimes q^{J_z}
\ee
The counit $\eps:\ \Uqsl\to\bbc$, as an algebra morphism, is defined by
\be
\eps(J_\pm)=0\ \ \ \ \ \eps(q^{\pm J_z})=1
\ee
The antipode $S:\ \Uqsl\to\Uqsl^{op\ cop}$ \footnote{$H^{op}$ is the ``opposite'' Hopf algebra defined by reversing the multiplication of $H$, while $H^{cop}$ is the ``coopposite'' Hopf algebra defined by reversing the comultipliction of $H$.}, as a bialgebra morphism, is defined by
\be
S(J_+)=-q^{+1}J_+\ \ \ \ S(J_-)=-q^{-1}J_-\ \ \ \ S(q^{\pm J_z})=q^{\mp J_z}
\ee

If $q=e^{-\o}$ is real, the Hopf $\star$-algebra $\Uqsu$ is defined by the Hopf algebra $\Uqsl$ implemented with a $\star$-structure
\be
(q^{J_z})^\star=q^{J_z}\ \ \ \ \ J_{\pm}^\star=J_{\mp}%q^{\mp1}
\ee
As $q\to1$, the quantum groups $\Uqsl$ and $\Uqsu$ recover the universal enveloping algebras $\text{U}(\fs\fl_2)$ and $\text{U}(\su)$ \cite{kassel}.

%\end{Definition}

%One can immediately check from the definition of the antipode that for all $x\in\Uqsu$
%\be
%S^2(x)=q^{2J_z}xq^{-2J_z}
%\ee
The quantum group $\Uqsu$ is a ribbon quasi-triangular Hopf algebra. A quasi-triangular Hopf algebra $A$ has a extra structure which is a invertible element $R\in A\otimes A$ satisfying some conditions \cite{kassel}. $R$ is called a universal R-matrix. Let's write $R=\sum_i a_i\otimes b_i$, then the element $u=\sum_i S(b_i)a_i$ is invertible, its inverse $u^{-1}=\sum_i S^{-2}(b_i) a_i$ (assuming the antipode is invertible). Then we have for all $x\in A$
\be
S^2(x)=uxu^{-1}\ \ \ \ \ \Delta(u)=(u\otimes u)(R_{21}R)^{-1}=(R_{21}R)^{-1}(u\otimes u)
\ee
A quasi-triangular Hopf algebra $A$ is a ribbon Hopf $\star$-algebra, if and only if there exists an invertible central element $\theta$ such that
\be
\Delta(\theta)=(R_{21}R)^{-1}(\theta\otimes \theta)\ \ \ \ \eps(\theta)=1\ \ \ \ S(\theta)=\theta
\ee
A useful element $\mu$ is defined by $\mu=\theta^{-1}u$, it will be used to define the quantum trace, as will be seen later. For the quantum group $\Uqsu$, the universal R-matrix is given by\footnote{The R-matrix and the $\theta$ center are defined because $\Uqsu$ is defined as a topological Hopf algebra, see \cite{kassel} for precise discussion. }
\be
R=q^{2J_z\otimes J_z}e_{q^{-1}}^{(q-q^{-1})(q^{J_z}J_+\otimes J_-q^{-J_z})} \ \ \ \ \text{where}\ \ \ \ e_\a^z=\sum_{k=0}^\infty \a^{-\frac{k(k-1)}{2}}\frac{z^k}{[k]_\a!}
\ee
and the central element $\theta$ and the element $\mu$ is given by
\be
\theta=q^{C_q}\ \ \ \ \mu=q^{2J_z}
\ee
where $C_q$ is the \emph{quantum Casimir element} which will equal $-2K(K+1)$ on the unitary irreps of $\Uqsu$.

The unitary representations of $\Uqsu$ are completely reducible, and unitary irreps are completely classified by a couple $(\o,K)\in\{1.-1\}\times\frac{1}{2}\mathbb{N}$. The appearance of $\o$ comes from the existence of the automorphism $\t_\o$ of $\Uqsu$ defined by $\t_\o(q^{J_z})=\o q^{J_z}$, $\t_\o(J_+)=\o^2J_+$ and $\t_\o(J_-)=J_-$, which doesn't have classical counterpart. In the following we only consider the unitary irreps with $\o=1$. We denote by $Irr(\Uqsu)$ the set of all equivalent classes of unitary irreps with $\o=1$. Each of the unitary irreps $\pi^K\in Irr(\Uqsu)$ is labeled by a negative half-integer $K$ (a spin). The carrier space is denoted by $V_K$ which has dimension $2K+1$. The representation of $\Uqsu$ on $V_K$ is given concretely by
\be
\pi^K(q^{J_z})\ e^K_m=q^m\ e^K_m\ \ \ \ \ \pi^K(J_{\pm})\ e^K_m=\sqrt{[K\pm m+1][K\mp m]}\  e^K_{m\pm1} %q^{\mp1/2}
\ee
where $e^K_m$ $(m=-K,\cdots,K)$ is the canonical basis. Given $M\in End(V_K)$, the \emph{quantum trace} $\tr_q(M)$ is defined by $\tr_q(M)=\tr_{V_j}(\mu^{-1} M)$. The definition of quantum trace is important for the proper definition of quantum group characters. In particular the q-dimension is given by $[d_K]=\tr_q(1)=[2K+1]$.

We have the Clebsch-Gordan decomposition of the tensor product representations, as the case of classical SU(2) group
\be
\pi^I\otimes\pi^J=\bigotimes_{K=|I-J|}^{I+J}\pi^K
\ee
For any three unitary irreps $\pi^I,\pi^J,\pi^K$, we define the Clebsch-Gordan maps $\psi^{K}_{IJ}\in \text{Hom}(V_I\otimes V_J,V_K)$ and $\phi_K^{IJ}\in\text{Hom}(V_K, V_I\otimes V_J)$. We define a function $Y(I,J,K)$
\be
Y(I,J,K)&=&1\ \ \ \ \text{if}\ \ \ \ I+J-K,\ J+K-I,\ K+I-J\in \mathbb{Z}^+\nonumber\\
Y(I,J,K)&=&0\ \ \ \ \text{otherwise}\label{Y}
\ee
When $Y(I,J,K)=0$, we have $\psi^K_{IJ}=\phi^{IJ}_K=0$. When $Y(I,J,K)\neq0$, $\psi^K_{IJ},\phi^{IJ}_K$ are nonzero and defined by the quantum Clebsch-Gordan coefficients \cite{KR}:
\be
\phi^{IJ}_K(e^K_c)=\sum_{a,b}\left(
                               \begin{array}{cc}
                                 a & b  \\
                                 I & J \\
                               \end{array}
                             \Bigg|
                             \begin{array}{c}
                                 K   \\
                                 c \\
                               \end{array}
                             \right)e_a^I\otimes e^J_b\ \ \ \ \
\psi_{IJ}^K(e_a^I\otimes e^J_b)=\sum_c\left(\begin{array}{c}
                                 c   \\
                                 K \\
                               \end{array}
                             \Bigg|\begin{array}{cc}
                                 I & J  \\
                                 a & b \\
                               \end{array}\right)e^K_c
\ee
The properties of the quantum Clebsch-Gordan coefficients is summarized in \cite{KR,roche}.

We define the linear forms $u^K_{ab}=\lag e_a^K|\pi^K(\cdot)|e_b^K\rag$, from which we obtain a Hopf $\star$-algebra $\Pol(\Uqsu)$ of the polynomial functions on the quantum group $\Uqsu$:

%\begin{Definition}

The polynomial algebra $\mathrm{Pol}(\Uqsu)$ over $\Uqsu$ is a Hopf $\star$-algebra linearly spanned by $\{u^K_{ab}\}_{I\in\frac{\mathbb{N}}{2};a,b=-I\cdots I}$, with the following algebraic relations
\be
&&u^I_1u^J_2=\sum_K\phi^{IJ}_Ku^K\psi^K_{IJ}\ \ \ \ \text{which implies}\ \ \ \ R^{IJ}_{12}u_1^Iu_2^J=u_2^Ju_1^I R^{IJ}_{12}\nonumber\\
&&\Delta(u^K_{ab})=\sum_{c}u^K_{ac}\otimes u^K_{cb}\nonumber\\
&&\eps(u_{ab}^K)=\delta_{ab}\ \ \ \ \ \eta(1)=u^0\nonumber\\
&&S(u^K_{ab})=\sum_{c,d}w^K_{bc}\ u_{cd}^K\ w^{K\ -1}_{da}\ \ \ \ \text{where}\ \ \ \ w^K_{ab}=\delta_{a,-b}q^{a}(-1)^{K-a}\nonumber\\
&&(u^K_{ab})^\star=S(u^K_{ba})\ \ \ \ \text{i.e.}\ \ \ \ u^K (u^K)^\dagger=(u^K)^\dagger u^K =1
\ee
All the above algebraic relations realize that $\mathrm{Pol}(\Uqsu)$ is the dual Hopf $\star$-algebra of $\Uqsu$.
%\end{Definition}

$\Pol(\Uqsu)$ can equivalently be defined by the enveloping algebra of the matrix elements of $u^{1/2}$, with the above algebraic relations. More explicitly, the quantum group $\slq$ is the Hopf algebra generated by the elements of\footnote{In the case that $q$ is a $d$-root of unity, there are additional relations that $a^{2d}=1$. etc. So in this case $\slq$ is a finite dimensional Hopf algebra.}
\be
u^{1/2}=\left(\begin{array}{cc}
a & b  \\
c & d \\
\end{array}\right)
\ee
satisfying the relations
\be
&&qab=ba\ \ \ \ qac=ca\ \ \ \ qbd=db\ \ \ \ qcd=dc\nonumber\\
&&bc=cb\ \ \ \ ad-da=(q^{-1}-q)bc\ \ \ \ ad-q^{-1}bc=1\nonumber\\
&&\Delta(a)=a\otimes a+b\otimes c\ \ \ \ \Delta(b)=b\otimes d+a\otimes b\ \ \ \ \Delta(c)=c\otimes a+d\otimes c\ \ \ \ \Delta(d)=d\otimes d+c\otimes b\nonumber\\
&&S(a)=d\ \ \ \ S(b)=-qb \ \ \ \ S(c)=-q^{-1}c\ \ \ \ S(d)=a
\ee
The matrix elements $u^I_{ab}$ are polynomials of the generators $(a,b,c,d)$ and form a basis of $\slq$ at lease for the case $q$ is not a root of unity (see \cite{klimyk} for the expression for $u^I_{ab}$).

If we restrict ourself with $q\in\mathbb{R}$, the Hopf $\star$-algebra $\mathrm{Pol}(\Uqsu)$ is given by by the Hopf algebra $\slq$ equipped with the $\star$-structure
\be
a^\star=d\ \ \ \ d^*=a\ \ \ \ b^\star=-q^{-1}c\ \ \ \ c^*=-qb
\ee
In addition, we can define a norm on $\Pol(\Uqsu)$ by
\be
||x||:=\sup_{\pi}||\pi(x)||\ \ \ \ \ \forall\ x\in\Pol(\Uqsu)
\ee
where the supremum is taken over all unitary representations of $\Pol(\Uqsu)$. After the completion of $\Pol(\Uqsu)$ with this norm, we obtain a unital $C^\star$-algebra which contains the Hopf $\star$-algebra $\Pol(\Uqsu)$ as a dense domain. We denote the resulting $C^\star$-algebra by $\suq$, which is an example of compact matrix quantum group defined by Woronowicz \cite{compactquantumgroup}.

Given a Hopf algebra $\ca$, a right invariant integral $h_R$ (or, left invariant integral $h_L$) of $\ca$ is a element in the dual Hopf algebra $\ca^*$, satisfying
\be
(h_R\otimes \id)\Delta=\eta\circ h_R\ \ \ \ &\text{or}&\ \ \ \ (\id\otimes h_L)\Delta=\eta\circ h_L
%h_R(a^\star)=\overline{h_R(a)}\ \ \ \ &\text{or}&\ \ \ \ h_L(a^\star)=\overline{h_L(a)}\ \ \ \ \ \  \forall\ a\in\ca
\ee
On the quantum group $\slq$, there exists a unique left \emph{and} right invariant integral, namely a Haar integral $h_{\slq}$, satisfying $h_{\slq}(1)=1$ \cite{klimyk,compactquantumgroup}, it is defined explicitly by
\be
&&h_{\slq}\big(a^\a b^\b c^\g d^\delta\big)\neq0\ \ \ \ \ \text{only if}\ \ \ \ \ \a=\delta\ \ \ \ \text{and}\ \ \ \ \b=\g\nonumber\\
&&h_{\slq}\big((bc)^n\big)=(-q)^n\frac{1-q^2}{1-q^{2(n+1)}}
\ee
For $q\in\mathbb{R}$, $h_{\slq}$ gives the Haar integral of $\suq$, which additionally satisfies
\be
h_{\slq}(a^\star)=\overline{h_{\slq}(a)}\ \ \ \ \ \  \forall\ a\in\suq.
\ee
and have similar property $h_{\slq}(u^I_{ab})=\delta_{I,0}$ as classical Haar integral.

\subsection{Comodule Algebra and Quantum Plane}

As a preparation for studying the coherent states of $\slq$, we review briefly the concepts of comodule algebra and the coaction of $\Uqsu$ on the quantum plane, see \cite{kassel} for a detailed discussion.

We first recall that a module\footnote{An $A$-module is also called a representation space of $A$ in representation theory.} of a algebra $A$ (a $A$-module) is a vector space $V$ together with a bilinear map $(a,v)\mapsto av$ from $A\times V\to V$ such that $a(a'v)=(aa')v$ (associativity) and $1v=v$ (unit) for all $a,a'\in A$ and $v\in V$. A tensor product of two $A$-modules $U\otimes V$ is a $A\otimes A$-module by $(a\otimes a')(u\otimes v)=au\otimes a'v$ for all $a,a'\in A$, $u\in U$ and $v\in V$. However if $A$ possesses a bialgebra structure with the comultiplication $\Delta$ and counit $\eps$, then $U\otimes V$ is naturally a $A$-module by $a(u\otimes v):=\Delta a(u\otimes v)=\sum_{(a)}a_{(1)}u\otimes a_{(2)}v$ (in Sweedler's sigma notation), and the counit equips any vector space $V$ with a trivial $A$-module structure $av=\eps(a)v$.

As algebras have action on modules, coalgebras have \emph{coaction} on \emph{comodules}. Given a coalgebra $(C,\Delta,\eps)$, a comodule is a vector space $N$ equipped with a linear map $\Delta_N:\ N\to C\otimes N$, called the coaction of $C$ on $N$, such that it satisfies
\be
(\Delta\otimes\id)\Delta_N=(\id\otimes\Delta_N)\Delta_N& \ \ \ \ \ \ &(\text{Coassociativity})\nonumber\\
\id=(\eps\otimes\id)\Delta_N\ \ \ \,& \ \ \ \ \ \ &(\text{counit}).
\ee
Given $(N,\Delta_N)$ and $(M,\Delta_M)$ be two $C$-comodules, a linear map $f$ from $N$ to $M$ is a morphism of $C$-comodules if
\be
(\id\otimes f)\circ\Delta_{N}=\Delta_{M}\circ f.
\ee
Given $(H,m,\eta,\Delta,\eps)$ a bialgebra and $M,N$ two $H$-comodules, we define a linear map $\Delta_{M\otimes N}$ by ($\t_{M,H}$ is a flip from $M\otimes H$ to $H\otimes M$)
\be
\Delta_{M\otimes N}:=(m\otimes\id_{M\otimes N})(\id_H\otimes \t_{M,H}\otimes\id_{N})(\Delta_M\otimes\Delta_N)
\ee
which endows the tensor product $M\otimes N$ an $H$-comodule structure.

Given a coalgebra $C$, its dual space $C^*$ then has naturally the algebra structure and vise versa, because of the dualities $<x,m(x^*\otimes y^*)>:=<\Delta x,(x^*\otimes y^*)>$ and $<x,1>:=\eps(x)$ where $x\in C$ and $x^*,y^*\in C^*$. Given $(N,\Delta_N)$ a $C$-comodule, its dual vector space $N^*$ can be naturally equipped with a $C^*$-module structure and vise versa. This duality is defined by
\be
<v,\ (x^*v^*)>:=<\Delta_N v,(x^*\otimes v^*)>
\ee

Given $(H,m,\eta,\Delta,\eps)$ a bialgebra and $(A,m_A,\eta_A)$ an algebra, we say $A$ is an $H$-comodule-algebra if (1) $A$ as a vector space has an $H$-comodule structure, given by a coaction $\Delta_A:\ A\to H\otimes A$, and (2) the structure maps $m_A:\ A\otimes A\to A$ and $\eta_A:\ \bbc\to A$ are morphisms of $H$-comodules, or equivalently, the coaction $\Delta_A$ is a morphism of algebras.

The quantum plane is an example of comodule algebra over the Hopf algebra of $\slq$ (with $q\in\bbc$ arbitrary complex number). First of all, a quantum plane $\bbc_q[x,y]$ is defined by the polynomial algebra generated by two variables $x$ and $y$, subjected to the nontrivial commutation relation
\be
yx=qxy
\ee
An $\slq$-comodule-algebra structure can be defined on the quantum plane $\bbc_q[x,y]\equiv A$ by the coaction $\Delta_A$, which can be expressed by the $\slq$ generators $a,b,c,d$
\be
\Delta_A(x):=a\otimes x+b\otimes y\ \ \ \ \ \ \ \Delta_A(y):=c\otimes x+d\otimes y
\ee
We can also rewrite these formulas in the matrix form
\be
\Delta_A\lt(\begin{array}{c}
    x  \\
    y  \\
  \end{array}\rt):=\lt(\begin{array}{cc}
    a & b \\
    c & d \\
  \end{array}\rt)\otimes\lt(\begin{array}{c}
    x  \\
    y  \\
  \end{array}\rt)
\ee
Note that the above definition of the coaction $\Delta_A$ doesn't involve the deformation parameter $q$, thus also valid for the coaction of classical SL(2) on a classical plane $\bbc[x,y]$.

It is clear that the elements $x^iy^j$ ($i,j\geq0$) form a basis in the quantum plane $\bbc_q[x,y]$. For each $x^iy^j$ the coaction is given explicitly by (see e.g. \cite{kassel})
\be
\Delta_A(x^iy^j)=\sum_{r=0}^i\sum_{s=0}^jq^{(i-r)s}\lt(\begin{array}{c}
    i  \\
    r  \\
  \end{array}\rt)_{q^2}\lt(\begin{array}{c}
    j  \\
    s  \\
  \end{array}\rt)_{q^2}a^rb^{i-r}c^sd^{j-s}\otimes x^{r+s}y^{i+j-r-s}\label{coaction}
\ee
where
\be
(n)_q=\frac{q^n-1}{q-1}\ \ \ \ \ \ \ \ \ (n)_q!=(n)_q(n-1)_q\cdots(1)_q\ \ \ \ \ \ \ \ \ \lt(\begin{array}{c}
    n  \\
    k  \\
  \end{array}\rt)_{q}=\frac{(n)_q!}{(k)_q!(n-k)_q!}
\ee
In particular for $x^i$ and $y^j$
\be
\Delta_A(x^i)=(a\otimes x+b\otimes y)^i\ \ \ \ \ \ \ \ \ \ \Delta_A(y^j)=(c\otimes x+d\otimes y)^j
\ee
which can be seen immediately from the fact that $\Delta_A$ is a morphism of algebra.

We denote by $\bbc_q[x,y]_n$ the subspace of degree $n$ polynomials in $\bbc_q[x,y]$, thus $\bbc_q[x,y]$ is a direct sum of the subspaces $\bbc_q[x,y]_n$. From Eq.(\ref{coaction}) we find that each $\bbc_q[x,y]_n$ is a sub-$\slq$-comodule-algebra. From the duality between a $C$-comodule and a $C^*$-module, we know that as $\bbc_q[x,y]_n$ is a comodule over $\suq$, the dual space $\bbc_q[x,y]_n^*$ is a module over $\Uqsu$, which is isomorphic to the irrep space $V_I$ with $2I=n$. In particular, the highest weight vector $e^I_I\equiv|I,I\rangle$ can be defined via the duality between $\bbc_q[x,y]_n$ and $\bbc_q[x,y]_n^*$ by
\be
<x^ky^{2I-k},e^I_I>=\delta_{2I,k}.
\ee

\subsection{Quantum Group Coherent State}

In this section, we define and study the quantum group coherent states as a quantum group analog of Perelomov's generalized coherent state on SU(2) \cite{Perelomov}, in terms of the language of comodule-algebra and quantum plane. Recall that for classical SU(2), a coherent state is defined by a action of SU(2) on the highest weight vector
\be
|j,\mathbf{n}\rangle:=\pi^j(\mathbf{n})|j,j\rangle\ \ \ \ \ \ \mathbf{n}\in\text{SU(2)}
\ee
whose component on the basis vector $|j,m\rangle$ are a function on SU(2) by
\be
\Psi_{j,m}(\mathbf{n}):=\lag j,m|\pi^j(\mathbf{n})|j,j\rag\ \ \ \ \ \ \text{or}\ \ \ \ \ \ \ \Psi_{j,m}^*(\mathbf{n}):=\lag j,j|\pi^j(\mathbf{n}^{-1})|j,m\rag.
\ee

We can find a quantum group analog of $\Psi_{j,m}^*(\mathbf{n}^{-1})=\lag j,j|\pi^j(\mathbf{n})|j,m\rag$ as a polynomial function in $\slq$. First of all, we have the isomorphism between $\bbc_q[x,y]_{2I}^*$ and the irrep $V_I$, then $\langle I,I|\equiv \langle e^I_I|$ can be identify as the element in $\bbc_q[x,y]_{2I}$. Actually $\langle I,I|$ is identified with $x^{2I}$ because of the duality $<x^{2I},e^I_I>=1$ and $<x^{2I},e^I_m>=0\ (m\neq I)$. Therefore the quantum group analog of $\Psi_{j,m}^*(\mathbf{n}^{-1})=\lag j,j|\pi^j(\mathbf{n})|j,m\rag$ can be given by
\be
<x^{2I},\pi^I(\mathbf{n})e^I_m>=<\Delta_A(x^{2I}),(\mathbf{n}\otimes e^I_m)>\ \ \ \ \ \ \forall\ \mathbf{n}\in \Uqsl
\ee
So we make the following definition for a \emph{dual} coherent state:

\begin{Definition}

The dual coherent state of $\slq$ is defined by $\Delta_A(x^{2I})\in\slq\otimes\bbc_q[x,y]$, where
\be
\Delta_A(x^{2I})&=&(a\otimes x+b\otimes y)^{2I}\nonumber\\
&=&\sum_{m=-I}^{I}\lt(\begin{array}{c}
    2I  \\
    I+m  \\
  \end{array}\rt)_{q^2}a^{I+m}b^{I-m}\otimes x^{I+m}y^{I-m}
\ee
which is a quantum group analog of $\langle j,j|\pi^j(n)$. %If $q$ is a primitive $k$th-root of unity, we assume $k>4I$.

\end{Definition}

It is not hard to see that the function
\be
<\Delta_A(x^{2I}),e^I_m>=<x^{2I},\pi^I(\cdot)e^I_m>=\lag e^I_I|\pi^I(\cdot)| e^I_m \rag=u^I_{Im}(\cdot)
\ee
is nothing but a matrix element of the $\Uqsl$ irreducible representation. The matrix elements $u^I_{mn}\in\suq$ have explicit expressions \cite{klimyk,repsu2}:
\be
u^I_{mn}&=&\lt(\begin{array}{c}
    2I  \\
    I+n  \\
  \end{array}\rt)_{q^2}^{\half}\lt(\begin{array}{c}
    2I  \\
    I+m  \\
  \end{array}\rt)_{q^2}^{-\half}\times\nonumber\\
  &&\sum_{s=\max\{0,m-n\}}^{\min\{I-n,I+m\}}\ \ q^{(n-m+s)s}\lt(\begin{array}{c}
    I+n  \\
    I+m-s  \\
  \end{array}\rt)_{q^2}\lt(\begin{array}{c}
    I-n \\
    s \\
  \end{array}\rt)_{q^2}a^{I+m-s}b^{n-m+s}c^sd^{I-n-s}
\ee
In particular,
\be
u^I_{Im}=\lt(\begin{array}{c}
    2I  \\
    I+m  \\
  \end{array}\rt)_{q^2}^{\half}a^{I+m}b^{I-m}\ \ \ \ \ \ \text{and}\ \ \ \ \ \ \
u^I_{mI}=\lt(\begin{array}{c}
    2I  \\
    I+m  \\
  \end{array}\rt)_{q^2}^{\half}a^{I+m}c^{I-m}
\ee
where $u^I_{Im}$ gives the explicit expression of $<\Delta_A(x^{2I}),e^I_m>$. Then $u^I_{Im}(\cdot)$ is a quantum group analog of $\lag j, j|\pi^j(\cdot)|j,m\rag$ for classical SU(2). Therefore clearly the quantum group analog of $\lag j,\mathbf{n}|j,m\rag=\lag j, j|\pi^j(\mathbf{n}^{-1})|j,m\rag$ ($\mathbf{n}\in\text{SU(2)}$) is the following object in $\slq$
\be
S(u^I_{Im})=\lt(\begin{array}{c}
    2I  \\
    I+m  \\
  \end{array}\rt)_{q^2}^{\half}S(b)^{I-m}S(a)^{I+m}=\lt(\begin{array}{c}
    2I  \\
    I+m  \\
  \end{array}\rt)_{q^2}^{\half}(c^\star)^{I-m}(a^\star)^{I+m}=(u^I_{mI})^\star
\ee
where the second and third equalities hold for $\suq$ with $q\in\mathbb{R}$.

Now we give a definition of the coherent state of $\slq$\footnote{The coherent state defined here is slightly different from the one in \cite{suqcoherent}. For instance they have slightly different over-completeness relation. But both of them have the same classical limit as $q\to1$.} (with $q$ an arbitrary complex number)

\begin{Definition}

A $\slq$ coherent state of the irreducible representation $V_I$ is an element in $\slq\otimes V_I$, denoted by $|I,\bf{u}\rangle$ such that
\be
|I,\mathbf{u}\rangle:=\sum_{m=-I}^Iu^I_{mI}|I,m\rangle=\sum_{m=-I}^I\lt(\begin{array}{c}
    2I  \\
    I+m  \\
  \end{array}\rt)_{q^2}^{\half}a^{I+m}c^{I-m}|I,m\rangle\label{coherentstate}
\ee
where the label $\mathbf{u}$ means the quadruple $(a,b,c,d)$. On the other hand, there exists a unique (nondegenerat) scalar product on $V_I$ such that the highest weight vector $|I,I\rangle$ is normalized. This scalar product identify $V_I^*$ with $V_I$ and gives $\lag I,m|I,n\rag=\delta_{mn}$. So we define $\langle I,\mathbf{u}|\in \slq\otimes V_I^*$ by
\be
\langle I,\mathbf{u}|:=\sum_{m=-I}^IS(u^I_{Im})\langle I,m|.\label{coherentstate1}
\ee

\end{Definition}

\noindent
We have the following properties of the quantum group coherent state:

\begin{description}
\item[Classical limit:] If we take the classical limit $q\to1$, the matrix elements $(a,b,c,d)$ in SU(2) can be parametrized by the Euler angles $\phi_1\in[0,4\pi]$, $\theta\in[0,\pi]$, $\phi_2\in[0,2\pi]$
\be
\lt(\begin{array}{cc}
    a & b \\
    c & d \\
  \end{array}\rt)=\lt(\begin{array}{cc}
    \cos\frac{\theta}{2}e^{i(\phi_2+\phi_1)/2} &\  i\sin\frac{\theta}{2}e^{i(\phi_2-\phi_1)/2} \\
    i\sin\frac{\theta}{2}e^{-i(\phi_2-\phi_1)/2} &\  \cos\frac{\theta}{2}e^{-i(\phi_2+\phi_1)/2} \\
  \end{array}\rt)
\ee
The coherent state $|I,\bf{u}\rangle$ recovers the coherent state on classical SU(2)
\be
|I,\mathbf{u}\rangle\to \pi^I(\phi_1,\theta,\phi_2)|I,I\rangle= e^{iI\phi_1}\sum_{m=-I}^I\lt(\begin{array}{c}
    2I  \\
    I+m  \\
  \end{array}\rt)^\half\lt(\cos\frac{\theta}{2}\rt)^{I+m}\lt(i\sin\frac{\theta}{2}\rt)^{I-m}e^{im\phi_2}|I,m\rangle.
\ee

\item[Over-completeness relation:] From the definition Eqs.(\ref{coherentstate}) and (\ref{coherentstate1}), we have:
\be
h_{\slq}\lt(|I,\mathbf{u}\rangle\langle\mathbf{u},I|\rt)=\sum_{m,n=-I}^I|I,m\rangle\langle n,I|h_{\slq}(u^I_{mI}S(u^I_{In}))=\frac{q^{-2I}}{[2I+1]}
\ee
In the last step we use the othogonality relation \cite{klimyk,compactquantumgroup}
\be
h_{\slq}(u^I_{mk}S(u^J_{In}))=\delta_{IJ}\delta_{mn}\delta_{kl}\frac{q^{-2k}}{[2I+1]}
\ee
Therefore we obtain the over-completeness relation
\be
1_I={q^{-2I}}{[2I+1]}\ h_{\slq}\lt(|I,\mathbf{u}\rangle\langle\mathbf{u},I|\rt).
\ee

\item[Overlapping function:] A coherent state $|I,\mathbf{u}\rangle$ is an element in $\slq\otimes V_I$, thus  we define the overlapping function $\lag I,\mathbf{u}_1| I,\mathbf{u}_2\rag$ (assuming $\mathbf{u}_1$ and $\mathbf{u}_2$ are independent) as an element in $\slq\otimes\slq$. We can use the formulation of dual coherent state and write
\be
\lag I,\mathbf{u}_1| I,\mathbf{u}_2\rag(\mathbf{n}_1\otimes\mathbf{n}_2)&=&<x^{2I},\ \pi^I(S(\mathbf{n}_1))\pi^I(\mathbf{n}_2)e^I_I>=<(\Delta\otimes\id)\Delta_A(x^{2I}),\ (S(\mathbf{n}_1)\otimes\mathbf{n}_2\otimes e^I_I)>\nonumber\\
&=&<(S\otimes\id\otimes\id)(\Delta\otimes\id)\Delta_A(x^{2I}),\ (\mathbf{n}_1\otimes\mathbf{n}_2\otimes e^I_I)>
\ee
for all $\mathbf{n}_1\otimes\mathbf{n}_2\in\Uqsu\otimes\Uqsu$. Therefore we obtain the identity
\be
\lag I,\mathbf{u}_1| I,\mathbf{u}_2\rag=<(S\otimes\id\otimes\id)(\Delta\otimes\id)\Delta_A(x^{2I}),\ e^I_I>
\ee
Moreover since the coaction is a morphism of algebra, we have
\be
(S\otimes\id\otimes\id)(\Delta\otimes\id)\Delta_A(x^{2I})=\Big[S(a)\otimes a\otimes x+S(b)\otimes c\otimes x+S(a)\otimes b\otimes y+S(b)\otimes d\otimes y\Big]^{2I}
\ee
Therefore as a result
\be
\lag I,\mathbf{u}_1| I,\mathbf{u}_2\rag=\Big[S(a_1)\otimes a_2+S(b_1)\otimes c_2\Big]^{2I}=\Big[d_1\otimes a_2+(-q)b_1\otimes c_2\Big]^{2I}\label{overlap}
\ee
%This expression of overlapping function makes sense even without the $\star$-structure on $\slq$. Thus we can consider Eq.\Ref{overlap} as a definition of the overlapping function
%\be
%\lag I,\mathbf{u}_1| I,\mathbf{u}_2\rag\in\slq\otimes\slq
%\ee
%for $\slq$ coherent states with $q$ an arbitrary complex number.
The expression Eq.(\ref{overlap}) shows manifestly the factorization property of the coherent state, which is essentially resulted by the fact that the quantum plane $\bbc_q[x,y]$ is a comodule-algebra and the coaction $\Delta_A$ is a morphism of algebra.

\item[Normalization:] We consider the squared norm of coherent state $\lag I,\mathbf{u}| I,\mathbf{u}\rag$. Since the entries of both bra and ket are identical, the squared norm $\lag I,\mathbf{u}| I,\mathbf{u}\rag$ is considered as an element in $\slq$ (different from the overlapping function). The squared norm can be evaluated by the property of antipode:
\be
\lag I,\mathbf{u}| I,\mathbf{u}\rag=\sum_{m=-I}^IS(u_{Im}^I)\ u_{mI}^I=1
\ee
Another way to see the normalization is from Eq.(\ref{overlap}), by identifying $\mathbf{u}_1=\mathbf{u}_2$ and replacing the tensor product by the $\slq$ multiplication. Then $\lag I,\mathbf{u}| I,\mathbf{u}\rag=1$ follows from the unity of the quantum determinant.

\end{description}

\section{Quantum Group Coherent Intertwiner}

\subsection{Quantum Group Coherent Intertwiner and Q-Closure Condition}

The coherent intertwiner for SU(2) group \cite{LS} is widely used in the spinfoam formulation of LQG and its semiclassical analysis. Here in the context of quantum group $\slq$, a coherent intertwiner is defined by the following

\begin{Definition} $\ $%In the following definition of $\slq$ coherent intertwiner, we restrict ourselves in the case $q$ is not a root of unity:
\begin{itemize}
\item Given $n$ $\slq$ coherent states $|I_1,\mathbf{u}_1\rangle,\cdots,|I_n,\mathbf{u}_n\rangle$ belonging respectively to $\slq\otimes V_{I_1},\cdots,\slq\otimes V_{I_n}$, an $\slq$ coherent intertwiner $||\vec{I},\vec{\mathbf{u}}\ \rangle:\ V_{I_1}\otimes\cdots\otimes V_{I_n}\to \slq^{\otimes n}$ is defined by
\be
\langle v_1\otimes\cdots\otimes v_n||\vec{I},\vec{\mathbf{u}}\ \rangle&:=&h_{\slq}\Big(\lag v_1|\pi^{I_1}|I_1,\mathbf{u}_1\rag\otimes\cdots\otimes\lag v_n|\pi^{I_n}|I_n,\mathbf{u}_n\rag\Big)\nonumber\\
&=&\sum_{\{m_i\}}h_{\slq}\Big(\prod_{i=1}^n\lag v_i|\pi^{I_i}|I_i,m_i\rag\Big)\bigotimes_{i=1}^n\lag I_i,m_i|I_i,\mathbf{u}_i\rag
\ee
for all $v_i\in V_{I_i}$

\item We can identify $v_i\in V_{I_i}$ with a polynomial in the quantum plane $\bbc_q[x,y]_{2I_i}$. Without loss of generality, $v_i$ is identified with $x^{I+m}y^{I-m}$. Then the objects $\lag v|\pi^{I}|I,\mathbf{u}\rag$ in the above definition are given by
\be
<(\Delta\otimes\id)\Delta_A(x^{I+m}y^{I-m}),\ e^I_I>.
\ee
\end{itemize}
\end{Definition}

Given a polynomial function $u\in\slq$ and a element $x\in\Uqsl$, we define a right action $\Rhd$ and a left action $\Lhd$ of $x$ on $u$ by (in terms of Sweedler notation\footnote{Sweedler's notation is used for the comultiplication $\Delta x=\sum_{(x)}x_{(1)}\otimes x_{(2)}$ and $\Delta^{(n)} x=\sum_{(x)}x_{(1)}\otimes\cdots\otimes x_{(n)}$.})
\be
x\Rhd u:=\sum_{(u)}u_{(1)}\lag x,u_{(2)}\rag\ \ \ \ \ x\Lhd u:=\sum_{(u)}\lag x,u_{(1)}\rag u_{(2)}
\ee
where $\lag\ ,\ \rag:\ \Uqsl\times\slq\to\bbc$ is the duality bracket between $\Uqsl$ and $\slq$. The right/left action satisfies the following properties
\be
&1)&\ \ xx'\Rhd u=x\Rhd(x'\Rhd u)\ \ \ \ \ \ \ \ \ \ \ \ \ \ \ \ \  \ \ \ \ \ \ \ \ xx'\Lhd u=x'\Lhd(x\Lhd u)\nonumber\\
&2)&\ \ x\Rhd(u_1u_2)=\sum_{(x)}x_{(1)}\Rhd u_1\ x_{(2)}\Rhd u_2\ \ \ \ \ \ \ x\Lhd(u_1u_2)=\sum_{(x)}x_{(1)}\Lhd u_1\ x_{(2)}\Lhd u_2\label{derivation}
\ee
for all $x,x'\in\Uqsl$ and $u,u_1,u_2\in\slq$. An action satisfying the second property is called a \emph{derivation}. By the invariance of Haar integration, we have that
\be
h_{\slq}(x\Rhd u)=h_{\slq}(x\Lhd u)=\lag x,1\rag h_{\slq}(u)=\eps(x)\ h_{\slq}(u)\ \ \ \ \
\ee
Given $n$ polynomial functions $u_1,\cdots,u_n\in\slq$, by using the second property in Eq.(\ref{derivation}), we have
\be
&&\eps(x)\ h_{\slq}(u_1\cdots u_n)=h_{\slq}\Big(x\Rhd (u_1\cdots u_n)\Big)=h_{\slq}\Big(\sum_{(x)}x_{(1)}\Rhd u_1\cdots x_{(n)}\Rhd u_n\Big)\nonumber\\
&&=\sum_{(x),(u)}h_{\slq}\Big(u^{(1)}_1\cdots u_n^{(1)}\Big)\lag x_{(1)},u_1^{(2)}\rag\cdots\lag x_{(n)},u_n^{(2)}\rag\nonumber\\
&&\eps(x)\ h_{\slq}(u_1\cdots u_n)=h_{\slq}\Big(x\Lhd (u_1\cdots u_n)\Big)=h_{\slq}\Big(\sum_{(x)}x_{(1)}\Lhd u_1\cdots x_{(n)}\Lhd u_n\Big)\nonumber\\
&&=\sum_{(x),(u)}\lag x_{(1)},u_1^{(1)}\rag\cdots\lag x_{(n)},u_n^{(1)}\rag h_{\slq}\Big(u^{(2)}_1\cdots u_n^{(2)}\Big)\label{close0}
\ee
It turns out that the above relation generalizes the closure condition in SU(2) spin-networks (see e.g.\cite{RS}) to the context of quantum group. We consider the special case that $x=X_\pm,X_z$ where
\be
X_+=J_+ q^{-J_z},\ \ \ \ \ X_-=q^{-J_z}J_-,\ \ \ \ \ X_z=q^{-J_z}\frac{q^{J_z}-q^{-J_z}}{q-q^{-1}}
\ee
\begin{itemize}
\item For $X_+=J_+ q^{-J_z}$ and $X_-=q^{-J_z}J_-$, we first assume $n=4$
\be
\Delta^{(4)}(J_\pm)&=&(\Delta\otimes\Delta)\Delta(J_\pm)=\Delta q^{-J_z}\otimes\Delta J_\pm+\Delta J_\pm\otimes\Delta q^{J_z}\nonumber\\
&=&q^{-J_z}\otimes q^{-J_z}\otimes(q^{-J_z}\otimes J_\pm+J_\pm\otimes q^{J_z})+(q^{-J_z}\otimes J_\pm+J_\pm\otimes q^{J_z})\otimes q^{J_z}\otimes q^{J_z}\nonumber\\
&=&q^{-J_z}\otimes q^{-J_z}\otimes q^{-J_z}\otimes J_\pm+q^{-J_z}\otimes q^{-J_z}\otimes J_\pm\otimes q^{J_z}+q^{-J_z}\otimes J_\pm\otimes q^{J_z}\otimes q^{J_z}+\nonumber\\
&&+J_\pm\otimes q^{J_z}\otimes q^{J_z}\otimes q^{J_z}
\ee
It is not hard to generalize the result to general $n$
\be
\Delta^{(n)}(J_\pm)=\sum_{i=1}^n\lt(q^{-J_z}\rt)^{\otimes (i-1)}\otimes J_\pm\otimes \lt(q^{J_z}\rt)^{\otimes(n-i)}
\ee
Since $\Delta$ is a morphism of algebra, we have $\Delta(J_+ q^{-J_z})=\Delta(J_+)\Delta(q^{-J_z})$ and $\Delta(q^{-J_z}J_-)=\Delta(q^{-J_z})\Delta(J_-)$. Therefore
\be
\Delta^{(n)}(X_\pm)&=&\sum_{i=1}^n\lt(q^{-2J_z}\rt)^{\otimes (i-1)}\otimes X_\pm\otimes 1^{\otimes(n-i)}
\ee

\item For $X_z=q^{-J_z}(q^{J_z}-q^{-J_z})/(q-q^{-1})$, since we have
\be
\Delta^{(n)}\lt(\frac{q^{J_z}-q^{-J_z}}{q-q^{-1}}\rt)&=&\frac{\lt(q^{J_z}\rt)^{\otimes n}-\lt(q^{-J_z}\rt)^{\otimes n}}{q-q^{-1}}\nonumber\\
&=&\sum_{i=1}^n\lt(q^{-J_z}\rt)^{\otimes (i-1)}\otimes \lt(\frac{q^{J_z}-q^{-J_z}}{q-q^{-1}}\rt)\otimes \lt(q^{J_z}\rt)^{\otimes(n-i)}
\ee
then
\be
\Delta^{(n)}(X_z)&=&\sum_{i=1}^n\lt(q^{-2J_z}\rt)^{\otimes (i-1)}\otimes X_z\otimes 1^{\otimes(n-i)}
\ee

\end{itemize}
For $X_+=J_+ q^{-J_z}$, $X_-=q^{-J_z}J_-$ and $X_z=q^{-J_z}(q^{J_z}-q^{-J_z})/(q-q^{-1})$, we have
\be
\eps(X_\pm)=\eps(X_z)=0
\ee
We suppose the polynomial functions $u_i=u^{K_i}_{a_ib_i}$. By using the relation
\be
\Delta (u^{K_i}_{a_ib_i})=\sum_{c_i}u^{K_i}_{a_ic_i}\otimes u^{K_i}_{c_ib_i}
\ee
we obtain from Eq.(\ref{close0}) that (the indices $c_i$ are summed in the following relations)
\be
0&=&\sum_{i=1}^n u^{K_1}_{a_1c_1}(q^{-2J_z})\cdots u^{K_{i-1}}_{a_{i-1}c_{i-1}}(q^{-2J_z})\
u^{K_{i}}_{a_{i}c_{i}}(\vec{X})\ u^{K_{i+1}}_{a_{i+1}c_{i+1}}(1)\cdots u^{K_{n}}_{a_{n}c_{n}}(1)\
h_{\slq}\Big(u^{K_1}_{c_1b_1}\cdots u^{K_n}_{c_nb_n}\Big)\nonumber\\
0&=&\sum_{i=1}^n h_{\slq}\Big(u^{K_1}_{a_1c_1}\cdots u^{K_n}_{a_nc_n}\Big)\ u^{K_1}_{c_1b_1}(q^{-2J_z})\cdots u^{K_{i-1}}_{c_{i-1}b_{i-1}}(q^{-2J_z})\
u^{K_{i}}_{c_{i}b_{i}}(\vec{X})\ u^{K_{i+1}}_{c_{i+1}b_{i+1}}(1)\cdots u^{K_{n}}_{c_{n}b_{n}}(1)\label{close}
\ee
for $\vec{X}=(X_+,X_-,X_z)$. Under the limit $q\to1$, the above relation recovers the closure condition for classical Lie algebra
\be
\sum_{i=1}^n\vec{X}_i=0
\ee
represented on the integral $\int_{\text{SU(2)}}\rmd g\ u^{K_1}_{a_1c_1}(g)\cdots u^{K_n}_{a_nc_n}(g)$. Therefore we define a \emph{q-closure condition} by

\begin{Definition}
The quantum deformed closure condition (q-closure condition) is given by
\be
0=\Delta^{(n)}(\vec{X})=\sum_{i=1}^n\lt(q^{-2J_z}\rt)^{\otimes (i-1)}\otimes \vec{X} \otimes 1^{\otimes(n-i)}
\ee
imposed on a product of $\Uqsl$ irreps $V_{I_1}\otimes\cdots\otimes V_{I_n}$, for $X_+=J_+ q^{-J_z}$, $X_-=q^{-J_z}J_-$ and $X_z=q^{-J_z}(q^{J_z}-q^{-J_z})/(q-q^{-1})$. Under the limit $q\to1$, the q-closure condition recovers the usual closure condition for classical Lie algebra $0=\Delta_{classical}^{(n)}(\vec{X})=\sum_{i=1}^n 1^{\otimes (i-1)}\otimes \vec{X} \otimes 1^{\otimes(n-i)}$, for $\vec{X}=(J_+,J_-,J_z)$.
\end{Definition}

From Eq.(\ref{close}) it is not hard to see that the $\slq$ coherent intertwiner satisfies the q-closure condition
\be
\Delta^{(n)}(\vec{X})||\vec{I},\vec{\mathbf{u}}\ \rangle=0.
\ee

\subsection{Squared Norm of Coherent Intertwiner and High Spin Limit}

Now we consider the squared norm of a 4-valent $\slq$ coherent intertwiner. This squared norm belongs to the tensor product space $\slq^{\otimes 4}$:
\be
\langle\vec{I},\vec{\mathbf{u}}\ ||\vec{I},\vec{\mathbf{u}}\ \rangle&=&\sum_{\{c_i\}}\bigotimes_{i=1}^4\Big[\lag I_i,\mathbf{u}_i|I_i,a_i\rag h_{\slq}\Big(\prod_{i=1}^4u^{I_i}_{a_ic_i}\Big)\Big]\bigotimes_{i=1}^4\Big[h_{\slq}\Big(\prod_{i=1}^4u^{I_i}_{c_ib_i}\Big)\lag I_i,b_i|I_i,\mathbf{u}_i\rag\Big]\nonumber\\
&=&\bigotimes_{i=1}^4\lag I_i,\mathbf{u}_i|I_i,a_i\rag h_{\slq}\Big(\prod_{i=1}^4u^{I_i}_{a_ib_i}\Big)\bigotimes_{i=1}^4\lag I_i,b_i|I_i,\mathbf{u}_i\rag\nonumber\\
&=&h_{\slq}\lt(\prod_{i=1}^4\lag I_i,\mathbf{u}_i|\pi^{I_i}|I_i,\mathbf{u}_i\rag\rt)
\ee
where in the second step we have used the invariance of Haar integral. Each factor of $\lag I_i,\mathbf{u}_i|\pi^{I_i}|I_i,\mathbf{u}_i\rag$ in the integrand can be computed in the same way as we did for the overlapping function, i.e.
\be
\lag I_i,\mathbf{u}_i|\pi^{I_i}|I_i,\mathbf{u}_i\rag&=&<(S\otimes\id^{\otimes 3})\Delta_A^{(4)}(x^{2I_i}),\ e^{I_i}_{I_i}>\nonumber\\
&=&\lt[\big(1\ \ 0\big)\otimes\lt(\begin{array}{cc}
    S(a_{i}) & S(b_{i}) \\
    S(c_{i}) & S(d_{i}) \\
  \end{array}\rt)\otimes\lt(\begin{array}{cc}
    a & b \\
    c & d \\
  \end{array}\rt)\otimes\lt(\begin{array}{cc}
    a_{i} & b_{i} \\
    c_{i} & d_{i} \\
  \end{array}\rt)\otimes\lt(\begin{array}{c}
    1   \\
    0  \\
  \end{array}\rt)\rt]^{2I_i}\nonumber\\
&=&\lt[\big(1\ \ 0\big)\otimes\lt(\begin{array}{cc}
    d_{i} & (-q)b_{i} \\
    (-q^{-1})c_{i} & a_{i} \\
  \end{array}\rt)\otimes\lt(\begin{array}{cc}
    a & b \\
    c & d \\
  \end{array}\rt)\otimes\lt(\begin{array}{cc}
    a_{i} & b_{i} \\
    c_{i} & d_{i} \\
  \end{array}\rt)\otimes\lt(\begin{array}{c}
    1   \\
    0  \\
  \end{array}\rt)\rt]^{2I_i}\nonumber\\
&=&\Big[S(a_i)\otimes a\otimes a_i+S(b_i)\otimes c\otimes a_i+S(a_i)\otimes b\otimes c_i+S(b_i)\otimes d\otimes c_i\Big]^{2I_i}\nonumber\\
&=&\Big[d_i\otimes a\otimes a_i+(-q)b_i\otimes c\otimes a_i+d_i\otimes b\otimes c_i+(-q)b_i\otimes d\otimes c_i\Big]^{2I_i}
\ee
where $\Delta^{(4)}_A=(\Delta\otimes\id\otimes\id)(\Delta\otimes\id)\Delta_A$. As the case for the overlapping function, the expression of $\lag I_i,\mathbf{u}_i|\pi^{I_i}|I_i,\mathbf{u}_i\rag$ is factorized, due to the fact that the quantum plane $\bbc_q[x,y]$ is a comodule-algebra, whose coaction $\Delta_A$ is a morphism of algebra.

We write down the Haar integral:
\be
\langle\vec{I},\vec{\mathbf{u}}\ ||\vec{I},\vec{\mathbf{u}}\ \rangle=h_{\slq}\lt(\prod_{i=1}^4\Big[d_i\otimes a\otimes a_i+(-q)b_i\otimes c\otimes a_i+d_i\otimes b\otimes c_i+(-q)b_i\otimes d\otimes c_i\Big]^{2I_i}\rt)\label{norm}
\ee
There are a few explanation for the above formula: The above $\prod_{i=1}^4$ can be considered as a usual classical product between different $\mathbf{u}_i=(a_i,b_i,c_i,d_i)$ and $\mathbf{u}_j=(a_j,b_j,c_j,d_j)$, but on the other hand, it should be considered as a $\slq$ product for the variables $(a,b,c,d)$, which should be integrated afterward. After the integration, the tensor product $\otimes$ between the variables from the same $\mathbf{u}_i$ (e.g. $d_i$ and $a_i$ in the first term) should be replaced by the $\slq$ product between them, because they belong to the same copy of $\slq$ as we are computing the squared norm of the coherent intertwiner.

The aim of this section is to study the large spin asymptotics of the squared norm $\langle\vec{I},\vec{\mathbf{u}}\ ||\vec{I},\vec{\mathbf{u}}\ \rangle$. The large spin limit in the context of quantum group is slightly different from the case for the classical group. The difference is that we should not only scale the spins, i.e. $I_i$ are relpaced by $\l I_i$ with $\l\to\infty$, but also at the same time scale the deformation parameter, i.e. $q=e^{-\o}$ are replaced by $q_\l=e^{-\o/\l}$ with $\l\to\infty$, while keeping $I_i\o\ll1$. In \cite{MT}, such a large spin limit was taken to find the asymptotic behavior of quantum 6j symbol corresponds to a path integral quantization of 3-dimensional gravity with a cosmological constant.

The integrand in Eq.\Ref{norm} is understood as a quantum group deformation of the quantity $\prod_{i=1}^4\lag j_i,\mathbf{n}_i|h|j_i,\mathbf{n}_i\rag$ for classical SU(2), and this deformation is valid for arbitrary complex number $q$. Furthermore, we can write the integrand into an expression as $e^S$.
\be
&&\prod_{i=1}^4\Big[d_i\otimes a\otimes a_i+(-q)b_i\otimes c\otimes a_i+d_i\otimes b\otimes c_i+(-q)b_i\otimes d\otimes c_i\Big]^{2I_i}\nonumber\\
&=&\prod_{i=1}^4 e^{2I_i\log\big[d_i\otimes a\otimes a_i+(-q)b_i\otimes c\otimes a_i+d_i\otimes b\otimes c_i+(-q)b_i\otimes d\otimes c_i\big]}\nonumber\\
&=&e^{\sum_{i=1}^42I_i\log\big[d_i\otimes a\otimes a_i+(-q)b_i\otimes c\otimes a_i+d_i\otimes b\otimes c_i+(-q)b_i\otimes d\otimes c_i\big]+R_{BCH}}\label{integrand}
\ee
where the exponential and logarithm are understood in the same way as they are defined for matrices. $R_{BCH}$ denotes the terms obtained through the Baker-Campbell-Hausdorff formula. For example, for two noncommutative variables $X$ and $Y$
\be
R_{BCH}:=\log(e^Xe^Y)-(X+Y)=\half[X,Y]+\frac{1}{12}[X,[X,Y]]-\frac{1}{12}[Y,[X,Y]]+\cdots
\ee
We denote the quantity on the exponential in Eq.(\ref{integrand}) by
\be
S&:=&S_{cl}+R_{BCH}\nonumber\\
S_{cl}&:=&\sum_{i=1}^42I_i\log\Big[d_i\otimes a\otimes a_i+(-q)b_i\otimes c\otimes a_i+d_i\otimes b\otimes c_i+(-q)b_i\otimes d\otimes c_i\Big]
\ee
If we make the replacements $I_i\mapsto\l I_i$ and $q\mapsto q_\l=e^{-\o/\l}$ and take the limit $\l\to\infty$, $S_{cl}$ goes asymptotically to the corresponding quantity from a classical SU(2) group, i.e. as $\l\to\infty$
\be
S_{cl}\sim \l\sum_{i=1}^42I_i\lag\half,\half\rt|\mathbf{n}_i^{-1}\ h\ \mathbf{n}_i\lt|\half,\half\rag\label{Scl}
\ee
where $\mathbf{n}_i, h\in\text{SU(2)}$ and
\be
\mathbf{n}_i=\lt(\begin{array}{cc}
    a_i & b_i \\
    c_i & d_i \\
  \end{array}\rt)\ \ \ \ \ \ \ \ h=\lt(\begin{array}{cc}
    a & b \\
    c & d \\
  \end{array}\rt)
\ee
This expression Eq.(\ref{Scl}) are employed in \cite{LS} for the high spin asymptotic analysis for the classical SU(2) coherent intertwiner. For the additional quantum group contribution $R_{BCH}$, it involves the commutators and all higher-order commutators between different
\be
2I_i\log\Big[d_i\otimes a\otimes a_i+(-q)b_i\otimes c\otimes a_i+d_i\otimes b\otimes c_i+(-q)b_i\otimes d\otimes c_i\Big]\equiv \Fx_i
\ee
We first consider the commutator $[\Fx_i,\Fx_j]$. Firstly $[\Fx_i,\Fx_j]$ proportional to $I_iI_j$, and it also gives the quantum group corrections to $S_{cl}$, which should be a term proportional to $\o$ plus the higher order terms of $\o$. Therefore
\be
[\Fx_i,\Fx_j]=I_iI_j\Big[\o f_{ij}(a_i,b_i,c_i,d_i;a,b,c,d) +o(\o^2)\Big]
\ee
where $f_{ij}$ is a function of the variables $(a_i,b_i,c_i,d_i)$ and $(a,b,c,d)$. If we make the replacements $I_i\mapsto\l I_i$ and $q\mapsto q_\l=e^{-\o/\l}$ and take the limit $\l\to\infty$, for the leading asymptotic behavior of $[\Fx_i,\Fx_j]$
\be
[\Fx_i,\Fx_j]\sim\l\ \o\ I_iI_j\ f_{ij}(a_i,b_i,c_i,d_i;a,b,c,d)
\ee
where the variables $(a_i,b_i,c_i,d_i)$ and $(a,b,c,d)$ in $f_{ij}$ behave classically in this limit. Similarly for the second-order commutator $[\Fx_i,[\Fx_j,\Fx_k]]$, it is proportional to $I_iI_jI_k$ and also gives corrections to the order $\o^2$ and higher order terms of $\o$, i.e.
\be
[\Fx_i,[\Fx_j,\Fx_k]]=I_iI_jI_k\Big[\o^2 f_{ijk}(a_i,b_i,c_i,d_i;a,b,c,d) +o(\o^3)\Big]
\ee
Under the scaling $I_i\mapsto\l I_i$ and $q\mapsto q_\l=e^{-\o/\l}$ and taking the limit $\l\to\infty$
\be
[\Fx_i,[\Fx_j,\Fx_k]]\sim\l\ \o^2\ I_iI_jI_k\ f_{ijk}(a_i,b_i,c_i,d_i;a,b,c,d)
\ee
where the variables $(a_i,b_i,c_i,d_i)$ and $(a,b,c,d)$ in $f_{ijk}$ also behave classically in the limit $\l\to\infty$. All the higher-order commutators can be analyzed in the same way. We obtain that for a $(n-1)$-order commutator
\be
[\Fx_{i_1},[\Fx_{i_2},[\cdots, [\Fx_{i_{n-1}},\Fx_{i_n}]\cdots]\sim \l\ \o^{n-1}\ I_{i_1}\cdots I_{i_n}\ f_{i_1\cdots i_n}(a_i,b_i,c_i,d_i;a,b,c,d)
\ee
As a result, under the scaling $I_i\mapsto\l I_i$ and $q\mapsto q_\l=e^{-\o/\l}$ and taking the limit $\l\to\infty$, the quantum group correction $R_{BCH}$ behaves asymptotically as
\be
R_{BCH}\sim \l\sum_{n=2}^{\infty}\o^{n-1}\sum^4_{i_1\cdots i_n=1}I_{i_1}\cdots I_{i_n}\ f_{i_1\cdots i_n}(\mathbf{n}_i;h)
\ee
where we have redefine the functions $f_{i_1\cdots i_n}$ by absorbing the numerical factor in front of each terms from the Baker-Campbell-Hausdorff formula. Under the limit $\l\to\infty$, each $f_{i_1\cdots i_n}(\mathbf{n}_i;h)$ can be understood as a classical function in terms of classical SU(2) variables $\mathbf{n}_i$ and $h$. Moreover if we assume $I_i\o\ll1$, the asymptotic behavior of $R_{BCH}$ is dominated only by the single commutator terms
\be
R_{BCH}\sim \l\ \o\ \sum_{j,k=1}^4I_jI_k\ f_{jk}(\mathbf{n}_i;h)\equiv R_\o
\ee
To summarize, we obtain the high spin asymptotic behavior of $S$ by
\be
S\sim S_{\text{eff}}:= \l\sum_{i=1}^42I_i\lag\half,\half\rt|\mathbf{n}_i^{-1}\ h\ \mathbf{n}_i\lt|\half,\half\rag +\l\ \o \sum_{j,k=1}^4I_jI_k\ f_{jk}(\mathbf{n}_i;h)\label{actionS}
\ee
where $\mathbf{n}_i,h$ are the classical SU(2) variables. %Note that the derivation of Eq.(\ref{actionS}) doesn't involves the $\star$-structure of $\suq$, so Eq.(\ref{actionS}) is not only valid for $q=]0,1[$ but also valid for $q$ at root of unity, i.e. $\o=\frac{2\pi i}{k}$ with an odd integer $k>2$.

Recall Eq.(\ref{norm}) that the squared norm of coherent intertwiner is a Haar integration of the integrand $e^S$. But since the integrand $e^S$ now becomes a classical function of SU(2) in the high spin limit, we argue that the Haar integration $h_{\slq}$ recovers the classical Haar integration on SU(2) in this limit. The reason is the following: suppose we expand the integrand $e^S$, where $S$ is given by Eq.\Ref{actionS}, into power series of the matrix element of $h\in\text{SU(2)}$ (the matrix elements of $h$ are the integration variables). $e^S$ is asymptotically a classical function of SU(2) under the high spin limit means that permuting the matrix elements of $h$ doesn't affect the asymptotic behavior of $e^S$. Suppose the matrix elements of $h$ are $(a,b,c,d)$, the quantum Haar integration $h_{\slq}$ is a functional that
\be
&&h_{\slq}\big(a^\a b^\b c^\g d^\delta\big)\neq0\ \ \ \ \ \text{only if}\ \ \ \ \ \a=\delta\ \ \ \ \text{and}\ \ \ \ \b=\g\nonumber\\
&&h_{\slq}\big((bc)^n\big)=\frac{(-q)^n}{(n+1)_{q^2}}
\ee
Since in the high spin limit, we scale the deformation parameter $q\mapsto q_\l=e^{-\o/\l}$, as $\l\to\infty$ the quantum Haar integration $h_{\slq}$ recovers its classical limit as a Haar integration on SU(2).

In the end, the high spin limit of $\langle\vec{I},\vec{\mathbf{u}}\ ||\vec{I},\vec{\mathbf{u}}\ \rangle$ is given by
\be
&&\int_{\text{SU(2)}}\rmd h\ e^{S_{\text{eff}}}\ \ \ \ \ \ \ \text{as}\ \ \ \ \l\to\infty\nonumber\\
&&S_{\text{eff}}=S_{cl}+R_\o\nonumber\\
&&\ \ \ \ \  =\l \sum_{i=1}^42I_i\log\lag\half,\half\rt|\mathbf{n}_i^{-1}\ h\ \mathbf{n}_i\lt|\half,\half\rag +\l\ \o \sum_{j,k=1}^4I_jI_k\ f_{jk}(\mathbf{n}_i;h).
\ee

\subsection{Quantum Group Correction to Coherent Intertwiner}

The purpose of this subsection is to give a explicite expression for the quantum group correction $R_\o$ in $S_{\text{eff}}$. The contribution $R_\o$ comes from the commutator $[\Fx_i,\Fx_j]$ where
\be
\Fx_i&=& 2I_i\log\Big[d_i\otimes a\otimes a_i+(-q)b_i\otimes c\otimes a_i+d_i\otimes b\otimes c_i+(-q)b_i\otimes d\otimes c_i\Big]\nonumber\\
&\equiv& 2I_i\log X_i
\ee
where $X_i\equiv d_i\otimes a\otimes a_i+(-q)b_i\otimes c\otimes a_i+d_i\otimes b\otimes c_i+(-q)b_i\otimes d\otimes c_i$. Then we compute the commutator in terms of formal power expansion of $\log X_i$:
\be
[\Fx_i,\Fx_j]&=&4I_iI_j[\log X_i, \log X_j]\nonumber\\
&=&4I_iI_j\lt[(X_i-1)-\frac{1}{2}(X_i-1)^2+\frac{1}{3}(X_i-1)^3+\cdots,\ (X_j-1)-\frac{1}{2}(X_j-1)^2+\frac{1}{3}(X_j-1)^3+\cdots\rt]\nonumber\\
&=&4I_iI_j\lt(1-(X_i-1)+(X_i-1)^2+\cdots\rt)\lt[X_i,X_j\rt]\lt(1-(X_j-1)+(X_j-1)^2+\cdots\rt)\nonumber\\
&=&4I_iI_jX^{-1}_i\lt[X_i,X_j\rt]X_j^{-1}
\ee
where we use the expression of formal geometric series in the last step. As we have shown previously, the resulting quantity of the commutator should recover its classical conter-part in the high spin limit. After the computation of the above commutator, we can freely permute among $X_i^{-1}, X_j^{-1}, \lt[X_i,X_j\rt]$, and their commutators only gives subleading contributions. Each $X_i$ is treated as a classical functions on SU(2).  Moreover we can also assume the above commutator is a quantization of a certain Poisson bracket, then the above result is obtained as the first order of $\hbar$.

In the high spin limit,
\be
X_i\sim \lag\half,\half\rt|\mathbf{n}_i^{-1}\ h\ \mathbf{n}_i\lt|\half,\half\rag
\ee
So we only need to compute the commutator $[X_i,X_j]$. As the leading contribution:
\be
&&[X_i,X_j]\nonumber\\
&\sim&\lt[\big(d_i\otimes a\otimes a_i+b_i\otimes c\otimes a_i+d_i\otimes b\otimes c_i+b_i\otimes d\otimes c_i\big),\ \big(d_j\otimes a\otimes a_j+b_j\otimes c\otimes a_j+d_j\otimes b\otimes c_j+b_j\otimes d\otimes c_j\big)\rt]\nonumber\\
&=&\lt[d_i\otimes a\otimes a_i,\ b_j\otimes c\otimes a_j\rt]+[d_i\otimes a\otimes a_i,\ d_j\otimes b\otimes c_j]+[d_i\otimes a\otimes a_i,\ b_j\otimes d\otimes c_j]+\nonumber\\
&&+[b_i\otimes c\otimes a_i,\ d_j\otimes a\otimes a_j]+[b_i\otimes c\otimes a_i,\ d_j\otimes b\otimes c_j]+[b_i\otimes c\otimes a_i,\ b_j\otimes d\otimes c_j]+\nonumber\\
&&+[d_i\otimes b\otimes c_i,\ d_j\otimes a\otimes a_j]+[d_i\otimes b\otimes c_i,\ b_j\otimes c\otimes a_j]+[d_i\otimes b\otimes c_i,\ b_j\otimes d\otimes c_j]+\nonumber\\
&&+[b_i\otimes d\otimes c_i,\ d_j\otimes a\otimes a_j]+[b_i\otimes d\otimes c_i,\ b_j\otimes c\otimes a_j]+[b_i\otimes d\otimes c_i,\ d_j\otimes b\otimes c_j]\nonumber\\
&\sim&\o\big[ ac( d_ib_j- b_id_j)a_ia_j
+ab ( a_ic_j- c_ia_j)d_id_j
+2bc (d_ib_j a_ic_j- b_id_j c_ia_j)\nonumber\\
&&+cd (a_ic_j- c_ia_j)b_ib_j
+bd (d_ib_j -b_id_j )c_ic_j\big]
\ee
where in the first step we put $q\to1$ in the expression of $X_i$ by ignoring the subleading contributions, and in the last step we replace all the variables by classical SU(2) matrix elements $(a,b,c,d)$.

By using the Baker-Campbell-Hausdorff formula
\be
e^{\Fx_1}e^{\Fx_2}e^{\Fx_3}e^{\Fx_4}&=&e^{\Fx_1+\Fx_2+\half [\Fx_1.\Fx_2]+\cdots}e^{\Fx_3+\Fx_4+\half [\Fx_3.\Fx_4]+\cdots}\nonumber\\
&=&e^{\Fx_1+\Fx_2+\Fx_3+\Fx_4+\half\sum_{i<j} [\Fx_i.\Fx_j]+\cdots}
\ee
where we ignore the higher order commutators. Therefore we obtain the expression of the quantum group correction $R_\o$ in the high spin limit
\be
R_\o&=&2\l\o\sum_{i<j}\frac{I_iI_j}{\lag \mathbf{n}_i| h |\mathbf{n}_i\rag\lag \mathbf{n}_j| h |\mathbf{n}_j\rag}\Big[ac( d_ib_j- b_id_j)a_ia_j
+ab ( a_ic_j- c_ia_j)d_id_j+\nonumber\\
&&+2bc (d_ib_j a_ic_j- b_id_j c_ia_j)+cd (a_ic_j- c_ia_j)b_ib_j
+bd (d_ib_j -b_id_j )c_ic_j\Big]
\ee
where $|\mathbf{n}\rangle=\mathbf{n}|1/2,1/2\rangle$ and
\be
\mathbf{n}_i=\lt(\begin{array}{cc}
    a_i & b_i \\
    c_i & d_i \\
  \end{array}\rt)\ \ \ \ \ \ \ \ h=\lt(\begin{array}{cc}
    a & b \\
    c & d \\
  \end{array}\rt)
\ee
are classical SU(2) group elements. $\mathbf{n}_i$ are the SU(2) group elements rotating the direction $\hat{z}=(0,0,1)\in\mathbb{R}^3$ into a direction $\hat{n}_i\in S^2\subset\mathbb{R}^3$. If we parametrize the unit vector $\hat{n}_i$ and the SU(2) variable $\mathbf{n}_i$ by the complex coordinate $z,\bar{z}$ on the sphere $S^2$
\be
\mathbf{n}_i&=&\frac{1}{\sqrt{1+|z_i|^2}}\lt(\begin{array}{cc}
    1 & z_i \\
    -\bar{z}_i & 1 \\
  \end{array}\rt)\nonumber\\
\hat{n}_i&=&\frac{-1}{{1+|z_i|^2}}\big(z_i+\bar{z}_i,\ iz_i-i\bar{z}_i,\ |z_i|^2-1\big).
\ee
We can also parametrize the SU(2) group element $h$ by
\be
h=p_\eta+i\vec{p}\cdot\vec{\sig}=\eta\sqrt{1-p^2}+i\vec{p}\cdot\vec{\sig}
\ee
where $\sig_i$ are Pauli matrices and $\eta=\pm1$. In this parametrization $\text{SU(2)}\simeq S^3$ splits into two hemispheres $B_\pm$ corresponding to $\eta=\pm1$. The SU(2) Haar measure can be written by
\be
\int\rmd h=\frac{1}{2\pi^2}\sum_{\eta=\pm1}\int_{B_\eta}\frac{\rmd^3\vec{p}}{\sqrt{1-p^2}}
\ee
In terms of these parametrizations, the classical part $S_{cl}$ reads
\be
S_{cl}=\l\sum_{i=1}^42I_i\log(p_\eta+i\vec{p}\cdot\hat{n}_i)
\ee
and the quantum group correction reads
\be
R_\o&=&2\l\o\sum_{i<j}\frac{I_iI_j}{(p_\eta+i\vec{p}\cdot\hat{n}_i)(p_\eta+i\vec{p}\cdot\hat{n}_j)(1+|z_i|^2)(1+|z_j|^2)}\Bigg[(p_1+i p_2) (z_i-z_j)(p_3-i\eta
   \sqrt{1-p^2})+\nonumber\\
&&+\bar{z}_j \left[z_j ({p_1}+i{p_2}) (-i z_i\eta
   \sqrt{1-p^2}-{p_3} z_i+2 {p_1}-2 i {p_2})+({p_1}-i {p_2})
   ({p_3}-i \eta\sqrt{1-p^2})\right]+\nonumber\\
&&+\bar{z}_i \bigg\{({p_2}+i {p_1})
   \lt[|z_j|^2 (\eta\sqrt{1-p^2}-i
   {p_3})+\eta\sqrt{1-p^2}+i {p_3}\rt]-z_i \Big[\bar{z}_j ({p_1}-i
   {p_2}) ({p_3}+i\eta \sqrt{1-p^2})+\nonumber\\
&&+({p_1}+i {p_2}) (-i z_j
   \eta\sqrt{1-p^2}-{p_3} z_j+2 {p_1}-2 i
   {p_2})\Big]\bigg\}\Bigg]
\ee
The high spin asymptotic behavior for the squared norm of quantum group coherent intertwiner is determined by the stationary phase analysis of the integral
\be
\frac{1}{2\pi^2}\sum_{\eta=\pm1}\int_{B_\eta}\frac{\rmd^3\vec{p}}{\sqrt{1-p^2}}e^{S_{cl}+R_\o}
\ee
In the case of a classical SU(2) coherent intertwiner the stationary phase analysis is studied in \cite{LS}, where in the case of a nondegenerate configuration of $\hat{n}_i$, the saddle point of $S_{cl}$ gives a flat tetrahedron with $\vec{p}=0$ or $h=1$. The faces of tetrahedron have areas $I_i$ $(i=1,\cdots,4)$ and have normals $\hat{n}_i$ ($i=1,\cdots,4$) in $\mathbb{R}^3$. The quantum group correction $R_\o$ deforms the critical point of $S_{cl}$. The critical point of $S_{\text{eff}}=S_{cl}+R_\o$ is expected to imply the non-flatness i.e. $\vec{p}\neq0$ or $h\neq1$, which is expected to give a curved tetrahedron at least for some certain values of $\o$ (so far $\o$ is an arbituary complex number). A detailed stationary phase analysis of the above intergral will be reported eleswhere \cite{DingHan}.

%\subsection{Saddle Point Approximation}

\section{Quantum Group Spinfoam Vertex}

\subsection{Quantum Group Vertex Amplitude and High Spin Limit}

The same technique can be applied to the asymptotic analysis of the quantum group spinfoam vertex amplitude. First of all we recall that the Euclidean EPRL spinfoam vertex can be written as the follows, if we choose the Barbero-Immirzi parameter $\g<1$ (see. e.g. \cite{semiclassical} for details)
\be
A^{\g<1}_v(k_{ij},\mathbf{n}_{ij})=(-1)^\chi\int\prod_{i=1}^5\rmd g_i^+\rmd g_i^-\prod_{i<j}P_{ij}^{\g<1}(k_{ij},\mathbf{n}_{ij},\mathbf{n}_{ji},g_i,g_j)
\ee
where $P^{\g<1}_{ij}$ are the coherent propagator given by
\be
P_{ij}^{\g<1}(k_{ij},\mathbf{n}_{ij},\mathbf{n}_{ji},g_i,g_j)&=&\lag j^+_{ij},\mathbf{n}_{ij}|(g_i^+)^{-1}g_j^+| j^+_{ji},\mathbf{n}_{ji}\rag
\lag j^-_{ij},\mathbf{n}_{ij}|(g_i^-)^{-1}g_j^-| j^-_{ji},\mathbf{n}_{ji}\rag\nonumber\\
&&j^\pm_{ij}=j^\pm_{ji}=\frac{1\pm\g}{2}k_{ij}
\ee
here $|j_{ij}^\pm,\mathbf{n}_{ij}\rangle$ are the coherent states on classical SU(2) and $\mathbf{n}_{ij}$ and $g^\pm_i$ are the group elements in classical SU(2). Moreover the coherent propagator can be factorized into
\be
P_{ij}^{\g<1}(k_{ij},\mathbf{n}_{ij},\mathbf{n}_{ji},g_i,g_j)&=&\lag \half,\mathbf{n}_{ij}\rt|(g_i^+)^{-1}g_j^+\lt| \half,\mathbf{n}_{ji}\rag^{2j_{ij}^+}
\lag \half,\mathbf{n}_{ij}\rt|(g_i^-)^{-1}g_j^-\lt| \half,\mathbf{n}_{ji}\rag^{2j_{ij}^-}
\ee
Therefore we can write the vertex amplitude $A_v^{\g<1}$ into the form
\be
A^{\g<1}_v(k_{ij},\mathbf{n}_{ij})=\int\prod_{i=1}^5\rmd g_i^+\rmd g_i^- \ e^{S_{EPRL}^{\g<1}(k_{ij},\mathbf{n}_{ij},g_i)}\label{vertex}
\ee
where
\be
S_{EPRL}^{\g<1}(k_{ij},\mathbf{n}_{ij},g_i)=\sum_{i<j}\lt[2j_{ij}^+\log\lag \half,\mathbf{n}_{ij}\rt|(g_i^+)^{-1}g_j^+\lt| \half,\mathbf{n}_{ji}\rag+2j_{ij}^-\log
\lag \half,\mathbf{n}_{ij}\rt|(g_i^-)^{-1}g_j^-\lt| \half,\mathbf{n}_{ji}\rag\rt]
\ee
The expression Eq.\Ref{vertex} is the starting point of the asymptotic analysis for the EPRL spinfoam model in \cite{semiclassical}, where the asymptotic formula of $A^{\g<1}_v$ is given by the stationary phase approximation of the integral of Eq.\Ref{vertex}.

In order to define a quantum deformed spinfoam vertex amplitude suitable for the high spin asymptotic analysis, we can firstly define a quantum group version of the coherent propagator $P_{ij}^{\g<1}$. Thus we consider the quantum deformation of the object
\be
\lag j_{ij},\mathbf{n}_{ij}|g_i^{-1}g_j| j_{ji},\mathbf{n}_{ji}\rag=\lag j_{ij},j_{ij}|\mathbf{n}_{ij}^{-1}g_i^{-1}g_j\mathbf{n}_{ji}| j_{ij},j_{ij}\rag
\ee
We know from the previous discussions that it corresponds to the following quantity in the context of quantum group,
\be
&&<x^{2I_{ij}},\ S(\mathbf{n}_{ij})S(g_i)g_j\mathbf{n}_{ji}e^{I_{ij}}_{I_{ij}}>=<(S\otimes S\otimes\id^{\otimes 3})\Delta^{(5)}(x^{2I_{ij}}),\ (\mathbf{n}_{ij}\otimes g_i\otimes g_j\otimes\mathbf{n}_{ji}\otimes e^{I_{ij}}_{I_{ij}})>\nonumber\\
&&\forall\ \mathbf{n}_{ij},g_i,g_j,\mathbf{n}_{ji}\in\Uqsl
\ee
Therefore the quantum group version of $\lag j_{ij},j_{ij}|\mathbf{n}_{ij}^{-1}g_i^{-1}g_j\mathbf{n}_{ji}| j_{ij},j_{ij}\rag$ is an element in $\slq^{\otimes 4}$, given by
\be
&&<(S\otimes S\otimes\id^{\otimes 3})\Delta^{(5)}(x^{2I_{ij}}),\ e^{I_{ij}}_{I_{ij}}>\nonumber\\
&=&\lt[\Big(1\ \ 0\Big)\otimes\lt(\begin{array}{cc}
    S(a_{ij}) & S(b_{ij}) \\
    S(c_{ij}) & S(d_{ij}) \\
  \end{array}\rt)\otimes\lt(\begin{array}{cc}
    S(a_i) & S(b_i) \\
    S(c_i) & S(d_i) \\
  \end{array}\rt)\otimes\lt(\begin{array}{cc}
    a_j & b_j \\
    c_j & d_j \\
  \end{array}\rt)\otimes\lt(\begin{array}{cc}
    a_{ji} & b_{ji} \\
    c_{ji} & d_{ji} \\
  \end{array}\rt)\otimes\lt(\begin{array}{c}
    1   \\
    0  \\
  \end{array}\rt)\ \rt]^{2I_{ij}}\nonumber\\
&=&\lt[\Big(1\ \ 0\Big)\otimes\lt(\begin{array}{cc}
    d_{ij} & -qb_{ij} \\
    -q^{-1}c_{ij} & a_{ij} \\
  \end{array}\rt)\otimes\lt(\begin{array}{cc}
    d_{i} & -qb_{i} \\
    -q^{-1}c_{i} & a_{i} \\
  \end{array}\rt)\otimes\lt(\begin{array}{cc}
    a_j & b_j \\
    c_j & d_j \\
  \end{array}\rt)\otimes\lt(\begin{array}{cc}
    a_{ji} & b_{ji} \\
    c_{ji} & d_{ji} \\
  \end{array}\rt)\otimes\lt(\begin{array}{c}
    1   \\
    0  \\
  \end{array}\rt)\ \rt]^{2I_{ij}}\nonumber\\
&=&\Bigg[\Big(d_{ij}\otimes d_i+b_{ij}\otimes c_i\Big)\otimes\Big(a_j\otimes a_{ji}+b_j\otimes c_{ji}\Big)+(-q)\Big(d_{ij}\otimes b_i+b_{ij}\otimes a_i\Big)\otimes\Big(c_j\otimes a_{ji}+d_j\otimes c_{ji}\Big)\Bigg]^{2I_{ij}}
\ee
which has an factorized expression. Therefore we can also write it in a form like $e^S$
\be
&&<(S\otimes S\otimes\id^{\otimes 3})\Delta^{(5)}(x^{2I_{ij}}),\ e^{I_{ij}}_{I_{ij}}>\nonumber\\
&=&e^{2I_{ij}\log\Big[\big(d_{ij}\otimes d_i+b_{ij}\otimes c_i\big)\otimes\big(a_j\otimes a_{ji}+b_j\otimes c_{ji}\big)+(-q)\big(d_{ij}\otimes b_i+b_{ij}\otimes a_i\big)\otimes\big(c_j\otimes a_{ji}+d_j\otimes c_{ji}\big)\Big]}
\ee
We now define a quantum group coherent propagator $\cp^{\g<1}_{ij,\pm}$ by
\be
&&\cp^{\g<1}_{ij,\pm}(K_{ij},\mathbf{u}_{ij},\mathbf{u}_{ji},\mathbf{u}^\pm_{i},\mathbf{u}^\pm_{j})\nonumber\\
&:=&<(S\otimes S\otimes\id^{\otimes 3})\Delta_\pm^{(5)}(x^{2I^\pm_{ij}}),\ e^{I^\pm_{ij}}_{I^\pm_{ij}}>\nonumber\\
&=&e^{2I^\pm_{ij}\log\Big[\big(d_{ij}\otimes d^\pm_i+b_{ij}\otimes c^\pm_i\big)\otimes\big(a^\pm_j\otimes a_{ji}+b^\pm_j\otimes c_{ji}\big)+(-q)\big(d_{ij}\otimes b^\pm_i+b_{ij}\otimes a^\pm_i\big)\otimes\big(c^\pm_j\otimes a_{ji}+d_j^\pm\otimes c_{ji}\big)\Big]}\nonumber\\
%&&e^{2I^-_{ij}\log\Big[\big(d_{ij}\otimes d^-_i+b_{ij}\otimes c^-_i\big)\otimes\big(a^-_j\otimes a_{ji}+b^-_j\otimes c_{ji}\big)+(-q)\big(d_{ij}\otimes b^-_i+b_{ij}\otimes a^-_i\big)\otimes\big(c^-_j\otimes a_{ji}+d_j^-\otimes c_{ji}\big)\Big]}\nonumber\\
&&\ \ \ \ \text{with}\ \ \ \  I^{\pm}_{ij}=\frac{1\pm\g}{2}K_{ij}\label{P}
\ee
where $\mathbf{u}_{ij}$ denotes the quadruple $(a_{ij},b_{ij},c_{ij},d_{ij})$ and similar for $\mathbf{u}_{ji},\mathbf{u}_i,\mathbf{u}_j$, the multiplication of the two exponentials are non-commutative multiplication because it involves the multiplication between the variables from the same $\mathbf{u}_{ij}$ and between the variables from the same $\mathbf{u}_{ji}$. The quantum group spinfoam vertex amplitude $\ca^{\g<1}_v$ is then given by the integration
\be
\ca^{\g<1}_v:=(-1)^\chi h_{\slq_+}^{\otimes 5}\otimes h_{\slq_-}^{\otimes 5}\Big(\prod_{i<j}\cp^{\g<1}_{ij,+}\prod_{i<j}\cp^{\g<1}_{ij,-}\Big)\label{vertex}
\ee
where an ordering is assigned for the product $\prod_{i<j}$.

\begin{figure}[h]
\begin{center}
\includegraphics[width=11cm]{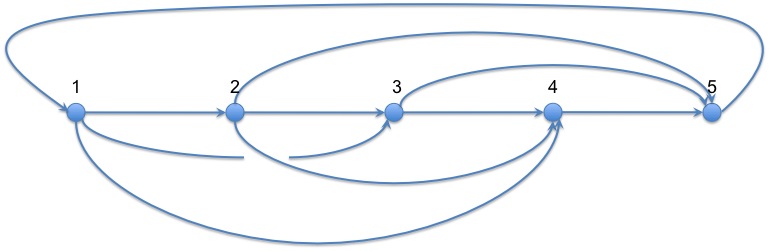}
\caption{The $\G_5^+$ graph.}
\label{gamma5}
\end{center}
\end{figure}

However the above definition of the vertex amplitude is heuristic since we ignore the universal $R$-matrix and the ribbon element $\mu^{-1}$ in the definition of $\cp^{\g<1}_{ij,\pm}$. In principle we should take them into account in order to define the vertex amplitude as an invariant of $\Uqsl$ representation\footnote{Here we assume $q$ is an arbitrary complex number since we define the model with $\slq$. But in the case of $q$ root of unity, in our opinion, the vertex amplitude here coincides with the Euclidean vertex amplitude proposed in \cite{EPRLq2}, once we implement the $R$-matrix and the ribbon element.}. Suppose we consider the vertex amplitude given by the following graph (the $\G_5^+$ in \cite{BCq,EPRLq}). In Fig.\ref{gamma5}, we order the nodes from left to right. Each oriented link $l_{ij}$ from node $i$ to node $j$ denotes a coherent propagator $\cp^{\g<1}_{ij,\pm}$. For the generic cases with $i<j$,
\be
\cp^{\g<1}_{ij,\pm}=\lag I^\pm_{ij},\mathbf{u}_{ij}\lt|S(\pi^{I^{\pm}_{ij}}_i)\pi^{I^\pm_{ij}}_j\rt|I^\pm_{ij},\mathbf{u}_{ji}\rag
\ee
which gives Eq.(\ref{P}). However for the link $l_{ij}$ from $i$ to $j$ with $i>j$, e.g. $\cp^{\g<1}_{51,\pm}$ in the case of $\G^+_5$, a ribbon element $\mu^{-1}$ should inserted in the coherent propagator
\be
\cp^{\g<1}_{51,\pm}=\lag I^\pm_{51},\mathbf{u}_{51}\lt|S(\pi^{I^{\pm}_{51}}_5)\pi^{I^\pm_{51}}(\mu^{-1})\pi^{I^\pm_{51}}_1\rt|I^\pm_{51},\mathbf{u}_{15}\rag
\ee
There is a crossing between $l_{13}$ and $l_{24}$ in $\G^+_5$. So a $R$-matrix should be inserted in their coherent propagators. If we write the $R$-matrix in the form $R=\sum_{R}R_1\otimes R_2$, then
\be
\cp^{\g<1}_{13,\pm}\cp^{\g<1}_{24,\pm}=\sum_{R}\lag I^\pm_{13},\mathbf{u}_{13}\lt|S(\pi^{I^{\pm}_{13}}_1)\pi^{I^\pm_{13}}(R_1)\pi^{I^\pm_{13}}_3\rt|I^\pm_{13},\mathbf{u}_{31}\rag
\lag I^\pm_{24},\mathbf{u}_{24}\lt|S(\pi^{I^{\pm}_{24}}_2)\pi^{I^\pm_{24}}(R_2)\pi^{I^\pm_{24}}_4\rt|I^\pm_{24},\mathbf{u}_{42}\rag
\ee
We can compute $\cp^{\g<1}_{51,\pm}$ and show that it has a factorized expression
\be
\cp^{\g<1}_{51}&=&<(S\otimes S\otimes\id^{\otimes 4})\Delta^{(6)}(x^{2I_{51}}),\ \cdots\otimes\cdots\otimes\mu^{-1}\otimes\cdots\otimes\cdots\otimes e^{I_{51}}_{I_{51}}>\nonumber\\
&=&\lt[\big(1\ \ 0\big)\otimes\lt(\begin{array}{cc}
    S(a_{51}) & S(b_{51}) \\
    S(c_{51}) & S(d_{51}) \\
  \end{array}\rt)\otimes\lt(\begin{array}{cc}
    S(a_5) & S(b_5) \\
    S(c_5) & S(d_5) \\
  \end{array}\rt)\otimes\lt(\begin{array}{cc}
    q^{-1} & 0 \\
    0 & q \\
  \end{array}\rt)\otimes\lt(\begin{array}{cc}
    a_1 & b_1 \\
    c_1 & d_1 \\
  \end{array}\rt)\otimes\lt(\begin{array}{cc}
    a_{15} & b_{15} \\
    c_{15} & d_{15} \\
  \end{array}\rt)\otimes\lt(\begin{array}{c}
    1   \\
    0  \\
  \end{array}\rt)\ \rt]^{2I_{51}}\nonumber\\
&=&e^{2I_{51}\log\lt[\big(1\ \ 0\big)\otimes\lt(\begin{array}{cc}
    S(a_{51}) & S(b_{51}) \\
    S(c_{51}) & S(d_{51}) \\
  \end{array}\rt)\otimes\lt(\begin{array}{cc}
    S(a_5) & S(b_5) \\
    S(c_5) & S(d_5) \\
  \end{array}\rt)\otimes\lt(\begin{array}{cc}
    q^{-1} & 0 \\
    0 & q \\
  \end{array}\rt)\otimes\lt(\begin{array}{cc}
    a_1 & b_1 \\
    c_1 & d_1 \\
  \end{array}\rt)\otimes\lt(\begin{array}{cc}
    a_{15} & b_{15} \\
    c_{15} & d_{15} \\
  \end{array}\rt)\otimes\lt(\begin{array}{c}
    1   \\
    0  \\
  \end{array}\rt)\ \rt]}
\ee
If we replace $q$ by $q_\l=e^{-\o/\l}$ and take the limit $\l\to\infty$
\be
\lt(\begin{array}{cc}
    q_\l^{-1} & 0 \\
    0 & q_\l \\
  \end{array}\rt)\to\lt(\begin{array}{cc}
    1\ & 0 \\
    0\ & 1 \\
  \end{array}\rt)
\ee
Therefore if we scale both the spins and the deformation parameter i.e. $I_{ij}\mapsto \l I_{ij}$ and $q\mapsto q_\l=e^{-\o/\l}$, and take the limit $\l\to\infty$
\be
\log\cp^{\g<1}_{51}\sim {2\l I_{51}\log\lt[\big(1\ \ 0\big)\otimes\lt(\begin{array}{cc}
    S(a_{51}) & S(b_{51}) \\
    S(c_{51}) & S(d_{51}) \\
  \end{array}\rt)\otimes\lt(\begin{array}{cc}
    S(a_5) & S(b_5) \\
    S(c_5) & S(d_5) \\
  \end{array}\rt)\otimes\lt(\begin{array}{cc}
    a_1 & b_1 \\
    c_1 & d_1 \\
  \end{array}\rt)\otimes\lt(\begin{array}{cc}
    a_{15} & b_{15} \\
    c_{15} & d_{15} \\
  \end{array}\rt)\otimes\lt(\begin{array}{c}
    1   \\
    0  \\
  \end{array}\rt)\ \rt]}
\ee
where the contribution of $\mu^{-1}$ is negligible in the high spin limit. Moreover $\mu^{-1}$ also doesn't contribute to the leading correction terms in $R_{BCH}$. As the previous analysis of $R_{BCH}$, the leading correction terms are proportional to $\o$ which comes from the commutators. So the leading contribution of $\mu^{-1}=1+o(\o)$ is $1$. As a result, the contribution from the ribbon element $\mu^{-1}$ can be ignored in the high spin asymptotic analysis. The same argument can also be applied to the coherent propagators with $R$-matrix, e.g. $\cp^{\g<1}_{13,\pm}\cp^{\g<1}_{24,\pm}$. Therefore as far as the asymptotic analysis is concerned, we can ignore the ribbon element $\mu^{-1}$ and the $R$-matrix in the definition of the vertex amplitude. The asymptotic behavior of the vertex amplitude Eq.(\ref{vertex}) is the same as the one with $\mu^{-1}$ and $R$-matrix implemented.

In the same way as the previous analysis for the squared norm of coherent intertwiner, we can express the integrand in Eq.\Ref{vertex} as
\be
\prod_{i<j}\cp^{\g<1}_{ij,+}\prod_{i<j}\cp^{\g<1}_{ij,-}=e^{\cs_{\g<0}}
\ee
where
\be
\cs_{\g<0}&=&\sum_{i<j}\sum_{\epsilon=\pm}2I^\epsilon_{ij}
\log\Big[\big(d_{ij}\otimes d^\epsilon_i+b_{ij}\otimes c^\epsilon_i\big)\otimes\big(a^\epsilon_j\otimes a_{ji}+b^\epsilon_j\otimes c_{ji}\big)+\nonumber\\
&&+(-q)\big(d_{ij}\otimes b^\epsilon_i+b_{ij}\otimes a^\epsilon_i\big)\otimes\big(c^\epsilon_j\otimes a_{ji}+d_j^\epsilon\otimes c_{ji}\big)\Big]+R_{BCH}\label{cs}
\ee
Here $R_{BCH}$ denotes the terms obtained through the Baker-Campbell-Hausdorff formula.

We now study the high spin limit of $\cs_{\g<0}$, by scaling both the spins and the deformation parameter $I_{ij}^\pm\mapsto\l I_{ij}^\pm$ and $q\mapsto q_\l=e^{-\o/\l}$, and taking the limit $\l\to\infty$, while keeping $I_{ij}^\pm\o\ll1$. Under this limit, the first term in Eq.(\ref{cs}) recovers the corresponding quantity for classical SU(2), which is nothing but the high spin limit of $S_{EPRL}^{\g<1}$
\be
S_{EPRL}^{\g<1}&\sim&\l\sum_{i<j}\lt[2I_{ij}^+\log\lag \half,\mathbf{n}_{ij}\rt|(g_i^+)^{-1}g_j^+\lt| \half,\mathbf{n}_{ji}\rag+2I_{ij}^-\log
\lag \half,\mathbf{n}_{ij}\rt|(g_i^-)^{-1}g_j^-\lt| \half,\mathbf{n}_{ji}\rag\rt]\nonumber\\
&&\ \ \ \ \text{with}\ \ \ \  I^{\pm}_{ij}=\frac{1\pm\g}{2}K_{ij}
\ee
The high spin limit of the quantum group corrections $R_{BCH}$ can be studied in the same way as we did for coherent intertwiner. $R_{BCH}$ involves the commutators and all higher-order commutators between
\be
\Fx_{ij}^\epsilon&\equiv& 2I^\epsilon_{ij}
\log\Big[\big(d_{ij}\otimes d^\epsilon_i+b_{ij}\otimes c^\epsilon_i\big)\otimes\big(a^\epsilon_j\otimes a_{ji}+b^\epsilon_j\otimes c_{ji}\big)+\nonumber\\
&&+(-q)\big(d_{ij}\otimes b^\epsilon_i+b_{ij}\otimes a^\epsilon_i\big)\otimes\big(c^\epsilon_j\otimes a_{ji}+d_j^\epsilon\otimes c_{ji}\big)\Big]
\ee
We first consider the commutator $[\Fx_{ij}^{\epsilon_1},\Fx_{kl}^{\epsilon_2}]$. Firstly $[\Fx_{ij}^{\epsilon_1},\Fx_{kl}^{\epsilon_2}]$ proportional to $I_{ij}^{\epsilon_1} I_{kl}^{\epsilon_2}$, and it should be a term proportional to $\o$ plus the higher order terms of $\o$. Therefore
\be
[\Fx_{ij}^{\epsilon_1},\Fx_{kl}^{\epsilon_2}]=I_{ij}^{\epsilon_1} I_{kl}^{\epsilon_2}\Big[\o \cf^{\epsilon_1\epsilon_2}_{ij,kl}(\mathbf{u}_{ij},\mathbf{u}_i) +o(\o^2)\Big]
\ee
If we make the replacements $I_{ij}^\pm\mapsto\l I_{ij}^\pm$ and $q\mapsto q_\l=e^{-\o/\l}$ and take the limit $\l\to\infty$, for the leading asymptotic behavior of $[\Fx_{ij}^{\epsilon_1},\Fx_{kl}^{\epsilon_2}]$
\be
[\Fx_{ij}^{\epsilon_1},\Fx_{kl}^{\epsilon_2}]\sim \l\ I_{ij}^{\epsilon_1} I_{kl}^{\epsilon_2}\ \o\ \cf^{\epsilon_1\epsilon_2}_{ij,kl}(\mathbf{n}_{ij},g_i)
\ee
where the function $\cf^{\epsilon_1\epsilon_2}_{ij,kl}$ behave classically in this limit, so we denote it by $\cf_{ij,kl}^{\epsilon_1\epsilon_2}(\mathbf{n}_{ij},g_i)$ in terms of classical SU(2) group element $\mathbf{n}_{ij},g_i$. Similarly for the second-order commutator $[\Fx_{ij}^{\epsilon_1},[\Fx_{kl}^{\epsilon_2},\Fx_{mn}^{\epsilon_3}]]$, it is proportional to $I_{ij}^{\epsilon_1} I_{kl}^{\epsilon_2}I_{mn}^{\epsilon_3}$ and also gives corrections to the order $\o^2$ and higher order terms of $\o$, i.e.
\be
[\Fx_{ij}^{\epsilon_1},[\Fx_{kl}^{\epsilon_2},\Fx_{mn}^{\epsilon_3}]]=I_{ij}^{\epsilon_1} I_{kl}^{\epsilon_2}I_{mn}^{\epsilon_3}\Big[\o^2 \cf^{\epsilon_1\epsilon_2\epsilon_3}_{ij,kl,mn}(\mathbf{u}_{ij},\mathbf{u}_i) +o(\o^3)\Big]
\ee
Under the scaling $I_{ij}^\pm\mapsto\l I_{ij}^\pm$ and $q\mapsto q_\l=e^{-\o/\l}$ and taking the limit $\l\to\infty$
\be
[\Fx_{ij}^{\epsilon_1},[\Fx_{kl}^{\epsilon_2},\Fx_{mn}^{\epsilon_3}]]\sim \l\ I_{ij}^{\epsilon_1} I_{kl}^{\epsilon_2}I_{mn}^{\epsilon_3}\ \o^2\ \cf^{\epsilon_1\epsilon_2\epsilon_3}_{ij,kl,mn}(\mathbf{n}_{ij},g_i)
\ee
where the function $\cf^{\epsilon_1\epsilon_2\epsilon_3}_{ij,kl,mn}$ also behave classically in the limit $\l\to\infty$. All the higher-order commutators can be analyzed in the same way. We obtain that for a $(n-1)$-order commutator
\be
[\Fx^{\epsilon_1}_{i_1j_1},[\Fx_{i_2j_2}^{\epsilon_2},[\cdots, [\Fx_{i_{n-1}j_{n-1}}^{\epsilon_{n-1}},\Fx_{i_nj_n}^{\epsilon_n}]\cdots]\sim \l\ \o^{n-1}\ I_{i_1j_1}^{\epsilon_1}\cdots I_{i_nj_n}^{\epsilon_n}\ \cf_{i_1j_1,\cdots, i_nj_n}^{\epsilon_1\cdots\epsilon_n}(\mathbf{n}_{ij},g_i)
\ee
As a result, under the high spin limit, the quantum group correction $R_{BCH}$ behaves asymptotically as
\be
R_{BCH}\sim \l\sum_{n=2}^{\infty}\o^{n-1}\sum_{i_1<j_1\cdots i_n<j_n}\sum_{\epsilon_1,\cdots,\epsilon_n=\pm}I_{i_1j_1}^{\epsilon_1}\cdots I_{i_nj_n}^{\epsilon_n}\ \cf^{\epsilon_1\cdots\epsilon_n}_{i_1j_1,\cdots, i_nj_n}(\mathbf{n}_{ij},g_i)
\ee
where we have redefine the functions $\cf^{\epsilon_1\cdots\epsilon_n}_{i_1j_1,\cdots, i_nj_n}$ by absorbing the numerical factor in front of each terms from the Baker-Campbell-Hausdorff formula. Under the limit $\l\to\infty$, each $\cf^{\epsilon_1\cdots\epsilon_n}_{i_1j_1,\cdots, i_nj_n}(\mathbf{n}_{ij},g_i)$ can be understood as a classical function on SU(2). Moreover if we assume $I_{ij}^\pm\o\ll1$, the asymptotic behavior of $R_{BCH}$ is dominated only by the single commutator terms
\be
R_{BCH}\sim \l\ \o\ \sum_{i<j,k<l}\sum_{\epsilon_1,\epsilon_2=\pm}I_{ij}^{\epsilon_1}I_{kl}^{\epsilon_2}\ \cf^{\epsilon_1\epsilon_2}_{ij,kl}(\mathbf{n}_{ij},g_i)
\ee
As a result the high spin asymptotic behavior of $\cs_{\g<1}$ defines an effective quantity $\cs_{\g<1,\text{eff}}$
\be
\cs_{\g<1}\sim\cs_{\g<1,\text{eff}}&:=& \l\sum_{i<j}\lt[2I_{ij}^+\log\lag \half,\mathbf{n}_{ij}\rt|(g_i^+)^{-1}g_j^+\lt| \half,\mathbf{n}_{ji}\rag+2I_{ij}^-\log
\lag \half,\mathbf{n}_{ij}\rt|(g_i^-)^{-1}g_j^-\lt| \half,\mathbf{n}_{ji}\rag\rt]+\nonumber\\
&&+\l\ \o\ \sum_{i<j,k<l}\sum_{\epsilon_1,\epsilon_2=\pm}I_{ij}^{\epsilon_1}I_{kl}^{\epsilon_2}\ \cf^{\epsilon_1\epsilon_2}_{ij,kl}(\mathbf{n}_{ij},g_i)\nonumber\\
&&\ \ \ \ \text{with}\ \ \ \  I^{\pm}_{ij}=\frac{1\pm\g}{2}K_{ij}\label{actionS1}
\ee
Under high spin limit, all the quantum group variables in $\cs_{\g<1}$ behavior classically. The resulting asymptotic behavior of $\cs_{\g<1}$ is a classical function over SU(2) variables, while all the quantum group corrections are given by the terms proportional to $\o$, i.e.
\be
\calr_\o:=\o\ \sum_{i<j,k<l}\sum_{\epsilon_1,\epsilon_2=\pm}I_{ij}^{\epsilon_1}I_{kl}^{\epsilon_2}\ \cf^{\epsilon_1\epsilon_2}_{ij,kl}(\mathbf{n}_{ij},g_i).
\ee

%\item The quantum group correction term proportional to the spin-square $I_{ij}^{\epsilon_1}I_{kl}^{\epsilon_2}$, i.e. it scales in the same way as a 4-volume, as we expected for the cosmological constant term of discrete GR. But further analysis should be carried out to determine the expression for the function $\cf^{\epsilon_1\epsilon_2}_{ij,kl}(\mathbf{n}_{ij},g_i)$, in order to compute the asymptotics of the vertex amplitude.

Furthermore in the definition Eq.(\ref{vertex}) of vertex amplitude $\ca_v^{\g<1}$, the Haar integration $h_{\slq}^{\otimes 5}$ recovers it classical limit as $\l\to\infty$, in the same reason as the previous argument for the squared norm of coherent intertwiner. Therefore in the high spin limit, the asymptotic behavior of quantum group vertex amplitude $\ca_v^{\g<1}$ is given by
\be
\ca_v^{\g<1}\sim\int\prod_{i=1}^5\rmd g_i^+\rmd g_i^-\ e^{\cs_{\g<1,\text{eff}}}=:\ca_{v,\text{eff}}^{\g<1}
\ee
where $\ca_{v,\text{eff}}^{\g<1}$ defines an effective vertex amplitude with quantum group corrections taken into account.

\subsection{Quantum Group Correction to Spinfoam Vertex}

The quantum group correction $\calr_\o$ for the vertex amplitude can be written in terms of the commutators via the Baker-Campbell-Hausdorff formula
\be
\calr_\o=\frac{1}{2}\sum_{(i,j)\prec(k,l)}[\Fx^\epsilon_{ij},\Fx^{\epsilon'}_{kl}]
\ee
note that for the definition of the vertex amplitude Eq.\Ref{vertex}, an order has been assigned for the pairs $(i,j)$. This correction can be computed explicitly in a similar way as we did for the coherent intertwiner. We denote
\be
X^\epsilon_{ij}&\equiv&\lt[\Big(1\ \ 0\Big)\otimes\lt(\begin{array}{cc}
    d_{ij} & -qb_{ij} \\
    -q^{-1}c_{ij} & a_{ij} \\
  \end{array}\rt)\otimes\lt(\begin{array}{cc}
    d_{i} & -qb_{i} \\
    -q^{-1}c_{i} & a_{i} \\
  \end{array}\rt)\otimes\lt(\begin{array}{cc}
    a_j & b_j \\
    c_j & d_j \\
  \end{array}\rt)\otimes\lt(\begin{array}{cc}
    a_{ji} & b_{ji} \\
    c_{ji} & d_{ji} \\
  \end{array}\rt)\otimes\lt(\begin{array}{c}
    1   \\
    0  \\
  \end{array}\rt)\ \rt]\nonumber\\
&=&\big(d_{ij}\otimes d^\epsilon_i+b_{ij}\otimes c^\epsilon_i\big)\otimes\big(a^\epsilon_j\otimes a_{ji}+b^\epsilon_j\otimes c_{ji}\big)+(-q)\big(d_{ij}\otimes b^\epsilon_i+b_{ij}\otimes a^\epsilon_i\big)\otimes\big(c^\epsilon_j\otimes a_{ji}+d_j^\epsilon\otimes c_{ji}\big)\nonumber\\
\Fx_{ij}^\epsilon&=& 2I^\epsilon_{ij}
\log X^\epsilon_{ij}
\ee
Then the commutator can be computed via
\be
[\Fx^\epsilon_{ij},\Fx^{\epsilon'}_{kl}]\sim\frac{1}{X^\epsilon_{ij}}[X^\epsilon_{ij},X^{\epsilon'}_{kl}]\frac{1}{X^{\epsilon'}_{kl}}
\ee
in the high spin limit. The high spin asymptotics of $X^\pm_{ij}$ is its classical group counter-part
\be
X^\pm_{ij}\sim \lag \half,\mathbf{n}_{ij}\rt|(g_i^\pm)^{-1}g_j^\pm\lt| \half,\mathbf{n}_{ji}\rag
\ee
For the commutator $[X^\epsilon_{ij},X^{\epsilon'}_{kl}]$, it is not hard to see that it vanishes if $\epsilon\neq\epsilon'$. When $\epsilon=\epsilon'$, with a tedious computation we obtain that ($*$ denotes the matrix element which does't affect the result)
\be
[X_{ij},X_{kl}]&\sim&\o\delta_{ik}\Bigg[
\Big(1\ \ 0\Big)\lt(\begin{array}{cc}
    d_{ij}b_{il} & b_{ij}d_{il} \\
    * & * \\
  \end{array}\rt)\lt(\begin{array}{cc}
    -c_id_{i} & -a_ib_{i} \\
    c_{i}d_i & a_{i}b_i \\
  \end{array}\rt)\lt(\begin{array}{cc}
    a_ja_l & b_jb_l \\
    c_jc_l & d_jd_l\\
  \end{array}\rt)\lt(\begin{array}{cc}
    a_{ji}a_{li} & * \\
    c_{ji}c_{li} & * \\
  \end{array}\rt)\lt(\begin{array}{c}
    1   \\
    0  \\
  \end{array}\rt)\nonumber\\
&&\ \ \ \ \ +\
\Big(1\ \ 0\Big)\lt(\begin{array}{cc}
    d_{ij}b_{il} & b_{ij}d_{il} \\
    * & * \\
  \end{array}\rt)\lt(\begin{array}{cc}
    -c_id_{i} & -a_ib_{i} \\
    c_{i}d_i & a_{i}b_i \\
  \end{array}\rt)\lt(\begin{array}{cc}
    a_jb_l & b_ja_l \\
    c_jd_l & d_jc_l\\
  \end{array}\rt)\lt(\begin{array}{cc}
    a_{ji}c_{li} & * \\
    c_{ji}a_{li} & * \\
  \end{array}\rt)\lt(\begin{array}{c}
    1   \\
    0  \\
  \end{array}\rt)\nonumber\\
&&\ \ \ \ \ +\
\Big(1\ \ 0\Big)\lt(\begin{array}{cc}
    d_{ij}d_{il} & b_{ij}d_{il} \\
    * & * \\
  \end{array}\rt)\lt(\begin{array}{cc}
    b_id_{i} \ \  & -b_ib_{i} \\
    0 & 2b_{i}c_i \\
  \end{array}\rt)\lt(\begin{array}{cc}
    a_jc_l & b_jd_l \\
    c_ja_l & d_jb_l\\
  \end{array}\rt)\lt(\begin{array}{cc}
    a_{ji}a_{li} & * \\
    c_{ji}c_{li} & * \\
  \end{array}\rt)\lt(\begin{array}{c}
    1   \\
    0  \\
  \end{array}\rt)\nonumber\\
&&\ \ \ \ \ +\
\Big(1\ \ 0\Big)\lt(\begin{array}{cc}
    d_{ij}d_{il} & b_{ij}b_{il} \\
    * & * \\
  \end{array}\rt)\lt(\begin{array}{cc}
    b_id_{i}\ \  & -b_id_{i} \\
    a_{i}c_i\ \  & -a_{i}c_i \\
  \end{array}\rt)\lt(\begin{array}{cc}
    a_jd_l & b_jc_l \\
    c_jb_l & d_ja_l\\
  \end{array}\rt)\lt(\begin{array}{cc}
    a_{ji}c_{li} & * \\
    c_{ji}a_{li} & * \\
  \end{array}\rt)\lt(\begin{array}{c}
    1   \\
    0  \\
  \end{array}\rt)\nonumber\\
&&\ \ \ \ \ +\
\Big(1\ \ 0\Big)\lt(\begin{array}{cc}
    -d_{ij}b_{il} & b_{ij}b_{il} \\
    * & * \\
  \end{array}\rt)\lt(\begin{array}{cc}
    -2b_ic_{i}   & 0 \\
    a_{i}c_i   & -a_{i}c_i \\
  \end{array}\rt)\lt(\begin{array}{cc}
    a_jc_l & b_jd_l \\
    c_ja_l & d_jb_l\\
  \end{array}\rt)\lt(\begin{array}{cc}
    a_{ji}a_{li} & * \\
    c_{ji}c_{li} & * \\
  \end{array}\rt)\lt(\begin{array}{c}
    1   \\
    0  \\
  \end{array}\rt)\nonumber\\
&&\ \ \ \ \ +\
\Big(1\ \ 0\Big)\lt(\begin{array}{cc}
    d_{ij}b_{il} & b_{ij}d_{il} \\
    * & * \\
  \end{array}\rt)\lt(\begin{array}{cc}
    2b_ic_{i} \  & 0 \\
    0 \  & -2b_{i}c_i \\
  \end{array}\rt)\lt(\begin{array}{cc}
    a_jd_l & b_jc_l \\
    c_jb_l & d_ja_l\\
  \end{array}\rt)\lt(\begin{array}{cc}
    a_{ji}c_{li} & * \\
    c_{ji}a_{li} & * \\
  \end{array}\rt)\lt(\begin{array}{c}
    1   \\
    0  \\
  \end{array}\rt)
\Bigg]\nonumber\\
&+&\o\delta_{jl}\Bigg[
\Big(1\ \ 0\Big)\lt(\begin{array}{cc}
    d_{ij}d_{kj} & b_{ij}b_{kj} \\
    * & * \\
  \end{array}\rt)\lt(\begin{array}{cc}
    d_id_{k}\ \  & b_ib_{k} \\
    c_{i}c_k\ \  & a_{i}a_k \\
  \end{array}\rt)\lt(\begin{array}{cc}
    a_jb_j & -a_jb_j \\
    c_jd_j & -c_jd_j\\
  \end{array}\rt)\lt(\begin{array}{cc}
    a_{ji}c_{jk} & * \\
    c_{ji}a_{jk} & * \\
  \end{array}\rt)\lt(\begin{array}{c}
    1   \\
    0  \\
  \end{array}\rt)\nonumber\\
&&\ \ \ \ \ +\
\Big(1\ \ 0\Big)\lt(\begin{array}{cc}
    d_{ij}b_{kj} & b_{ij}d_{kj} \\
    * & * \\
  \end{array}\rt)\lt(\begin{array}{cc}
    -d_ic_{k}\ \  & b_ia_{k} \\
    c_{i}d_k\ \  & a_{i}b_k \\
  \end{array}\rt)\lt(\begin{array}{cc}
    a_jb_j & -a_jb_j \\
    c_jd_j & -c_jd_j\\
  \end{array}\rt)\lt(\begin{array}{cc}
    a_{ji}c_{jk} & * \\
    c_{ji}a_{jk} & * \\
  \end{array}\rt)\lt(\begin{array}{c}
    1   \\
    0  \\
  \end{array}\rt)\nonumber\\
&&\ \ \ \ \ +\
\Big(1\ \ 0\Big)\lt(\begin{array}{cc}
    d_{ij}d_{kj} & b_{ij}d_{kj} \\
    * & * \\
  \end{array}\rt)\lt(\begin{array}{cc}
    -d_ib_{k}\ \  & -b_id_{k} \\
    c_{i}b_k\ \  & a_{i}d_k \\
  \end{array}\rt)\lt(\begin{array}{cc}
    a_jc_j & b_jd_j \\
    -a_jc_j & -b_jd_j\\
  \end{array}\rt)\lt(\begin{array}{cc}
    a_{ji}a_{jk} & * \\
    c_{ji}c_{jk} & * \\
  \end{array}\rt)\lt(\begin{array}{c}
    1   \\
    0  \\
  \end{array}\rt)\nonumber\\
&&\ \ \ \ \ +\
\Big(1\ \ 0\Big)\lt(\begin{array}{cc}
    d_{ij}d_{kj} & b_{ij}b_{kj} \\
    * & * \\
  \end{array}\rt)\lt(\begin{array}{cc}
    d_ib_{k}\ \  & b_id_{k} \\
    c_{i}a_k\ \  & a_{i}c_k \\
  \end{array}\rt)\lt(\begin{array}{cc}
    2b_jc_j\  & 0 \\
    0\  & -2b_jc_j\\
  \end{array}\rt)\lt(\begin{array}{cc}
    a_{ji}c_{jk} & * \\
    c_{ji}a_{jk} & * \\
  \end{array}\rt)\lt(\begin{array}{c}
    1   \\
    0  \\
  \end{array}\rt)\nonumber\\
&&\ \ \ \ \ +\
\Big(1\ \ 0\Big)\lt(\begin{array}{cc}
    -d_{ij}b_{kj} & b_{ij}b_{kj} \\
    * & * \\
  \end{array}\rt)\lt(\begin{array}{cc}
    d_ia_{k}\ \  & b_ic_{k} \\
    -c_{i}a_k\ \  & -a_{i}c_k \\
  \end{array}\rt)\lt(\begin{array}{cc}
    a_jc_j & b_jd_j \\
    -a_jc_j & -b_jd_j\\
  \end{array}\rt)\lt(\begin{array}{cc}
    a_{ji}a_{jk} & * \\
    c_{ji}c_{jk} & * \\
  \end{array}\rt)\lt(\begin{array}{c}
    1   \\
    0  \\
  \end{array}\rt)\nonumber\\
&&\ \ \ \ \ +\
\Big(1\ \ 0\Big)\lt(\begin{array}{cc}
    d_{ij}b_{kj} & b_{ij}d_{kj} \\
    * & * \\
  \end{array}\rt)\lt(\begin{array}{cc}
    d_ia_{k}\ \  & b_ic_{k} \\
    c_{i}b_k\ \  & a_{i}d_k \\
  \end{array}\rt)\lt(\begin{array}{cc}
    -2b_jc_j & 0 \\
    0 & 2b_jc_j\\
  \end{array}\rt)\lt(\begin{array}{cc}
    a_{ji}c_{jk} & * \\
    c_{ji}a_{jk} & * \\
  \end{array}\rt)\lt(\begin{array}{c}
    1   \\
    0  \\
  \end{array}\rt)
\Bigg]
\ee
where we ignore the $\pm$ for the matrix elements $(a^\pm_i,b^\pm_i,c^\pm_i,d^\pm_i)$, and
\be
\lt(\begin{array}{cc}
    a^\pm_i & b^\pm_i \\
    c^\pm_i & d^\pm_i \\
  \end{array}\rt)=g_{i}^\pm\ \ \ \ \ \ \text{and}\ \ \ \ \ \ \lt(\begin{array}{cc}
    a_{ij} & b_{ij} \\
    c_{ij} & d_{ij} \\
  \end{array}\rt)=\mathbf{n}_{ij}=\frac{1}{\sqrt{1+|z_{ij}|^2}}\lt(\begin{array}{cc}
    1 & z_{ij} \\
    -\bar{z}_{ij} & 1 \\
  \end{array}\rt)
\ee
are classical SU(2) group elements. $\mathbf{n}_{ij}$ are the SU(2) group elements rotating the direction $\hat{z}=(0,0,1)\in\mathbb{R}^3$ into a direction $\hat{n}_{ij}\in S^2\subset\mathbb{R}^3$. We have parametrized $\mathbf{n}_{ij}$ in terms of the complex coordinate $z_{ij},\bar{z}_{ij}$ on the sphere $S^2$.

\section{Conclusion}

In this work we analyze the high spin asymptotics for both the squared norm of coherent intertwiner and the Euclidean vertex amplitude in the context of quantum group. We express both of them in terms of Haar integrations over quantum group, and show that in the high spin limit, they can be written as a classical integration over classical group. Moreover we identify and compute the quantum group corrections to both coherent intertwiner and spinfoam vertex amplitude.

For both quantum group coherent intertwiner and vertex amplitude, the resulting $S_{\text{eff}}$ and $\cs_{\g<1,\text{eff}}$ are now ready for the stationary phase analysis. We should further carry out the stationary phase analysis to both of them, in order to see (1.) whether the high spin limit of coherent intertwiner corresponds to a curved tetrahedron, and (2.) whether the high spin limit of the vertex amplitude corresponds to a curved 4-simplex, whose curvature comes from a cosmological constant, and whether the vertex amplitude gives a Regge gravity with a cosmological constant as its high spin asymptotics. In addition, about the coherent intertwiner, it is also interesting to study its expectation value of the geometrical operators (e.g. area and angle), in order to understand in detail the effective classical geometry obtained from quantum group.

On the other hand, the technique developed in this article can be carried out also for the Lorentzian quantum group spinfoam model. The asymptotic analysis of the Lorentzian quantum group spinfoam vertex will be studied in \cite{DingHan}.

% all the computation I have done is actually for SL_q(2) hopf algebra without using *-structure. I can assume q is a generic complex number but not a root of unity. If q is a root of unity, the duality between SL_q(2) and U_q(sl_2) is not clear to me. If q is real, my computation is pretty safe.

% the main purpose of this article is to develop a new computation tool to study the asymptotics of quantum group spinfoam. In the article we can see how the noncommutative q-objects recover the usual commutative objects in the high spin limit and at the same time give the quantum group corrections to the classical commutative objects.

\section*{Acknowledgments}

The authors thank Eugenio Bianchi, Thomas Krajewski, Karim Noui, Carlo Rovelli, Wolfgang Wieland, and Mingyi Zhang for fruitful discussions.

\end{document}